\documentclass[fleqn,usenatbib]{mnras}

\usepackage{newtxtext,newtxmath}
\usepackage[flushleft]{threeparttable}
\usepackage{lineno}
\usepackage[T1]{fontenc}
\usepackage{url}
\usepackage{rotating}

\DeclareRobustCommand{\VAN}[3]{#2}
\let\VANthebibliography\thebibliography
\def\thebibliography{\DeclareRobustCommand{\VAN}[3]{##3}\VANthebibliography}

\usepackage{graphicx}	
\usepackage{amsmath}	
\usepackage{xcolor}

\usepackage{ulem}

\title[VLBI meets \textit{Fermi}-LAT]{The jet apparent motion and central engine study of \textit{Fermi} blazars}
\author[Xiao et al.]{
H. B. Xiao,$^{1}$\thanks{E-mail: hubing.xiao@shnu.edu.cn}
J. T. Zhu,$^{3,5}$\thanks{E-mail: jingtian.zhu@studenti.unipd.it}
J. H. Fan,$^{2,4,6}$\thanks{E-mail: fjh@gzhu.edu.cn}
Z. Y. Pei,$^{2,4,6}$
Z. J. Luo,$^{1}$
S. H. Zhang,$^{1}$
\\
$^{1}$Shanghai Key Lab for Astrophysics, Shanghai Normal University, Shanghai 200234, China\\
$^{2}$Center for Astrophysics, Guangzhou University, Guangzhou 510006, China\\
$^{3}$Department of Physics and Astronomy ``G. Galilei", University of Padova,
Padova PD 35131, Italy\\
$^{4}$Key Laboratory for Astronomical Observation and Technology of Guangzhou, Guangzhou University, Guangzhou 510006, China\\
$^{5}$Istituto Nazionale di Fisica Nucleare, Padova PD 35131, Italy\\
$^{6}$Astronomy Science and Technology Research Laboratory of Department of Education of Guangdong Province, Guangzhou 510006, China\\
}

\date{Accepted XXX. Received YYY; in original form ZZZ}

\pubyear{2022}

\begin{document}
\label{firstpage}
\pagerange{\pageref{firstpage}--\pageref{lastpage}}
\maketitle

\begin{abstract}
The study of blazar jet has been performed for several decades via VLBI technique, while its generation and propagation stay unclear.
In the present work, we compiled a sample of 407 VLBI detected \textit{Fermi} blazars (VFBs) and studied the correlations between apparent velocity (${\rm log}\,\beta_{\rm app}$) and jet/accretion disk properties.
We found a positive correlation between $\gamma$-ray luminosity (${\rm log}\,L_{\rm \gamma}$) and ${\rm log}\,\beta_{\rm app}$, the correlation suggests that the apparent motion of jet knot is related to the jet power.
The correlations between ${\rm log}\, \beta^{\rm max}_{\rm app}$ and the jet radiation power (${\rm log}\, L_{\rm rad}$) and between ${\rm log}\,\beta^{\rm max}_{\rm app}$ and the jet extended region luminosity at 5 GHz (${\rm log}\, L_{\rm 5GHz}^{\rm ext}$), which is an indicator of jet kinetic power, reveal that the knots apparent motion is correlated with both jet radiation power and the kinetic power.
But this indication is not held for FSRQs in terms of the correlation ${\rm log}\, L_{\rm 5GHz}^{\rm ext}\, vs \, {\rm log}\, \beta^{\rm max}_{\rm app}$.
Besides, ${\rm log}\, \beta^{\rm max}_{\rm app}$ has a moderate correlation with accretion disk luminosity (${\rm log}\, L_{\rm Disk}$) and the normalized accretion disk luminosity ${\rm log}\, (L_{\rm Disk}/L_{\rm Edd})$, this may suggest both the power of accretion disk and the accretion rate are critical to generate knots and to accelerate them.

In addition, we found the VFBs have larger average values of $\gamma$-ray luminosity (${\rm log}\, L_{\rm \gamma}$), $\gamma$-ray photon index ($\alpha_{\rm ph}$), and variability index (${\rm log}\, VI$) than the rest of \textit{Fermi} blazars.
Through \textit{Gaussian mixture models} method, we generated a criteria, ${\rm log}\,L_{\rm \gamma} > 45.40$, $\alpha_{\rm ph} > 2.24$ and ${\rm log}\,VI > 1.71$ to find VFB candidates, selected 228 VFB candidates from the rest of \textit{Fermi} blazar.

\end{abstract}




\section{Introduction}
Blazars, an extreme subclass of active galactic nuclei (AGNs), are classified as flat-spectrum radio quasars (FSRQs) and BL Lacertae objects (BL Lac).
The former is characterized by strong emission lines, while the latter shows featureless optical spectra or weak emission lines \citep{Scarpa1997}.
Blazars have a broad electromagnetic emission range (from radio to $\gamma$-ray bands) and demonstrate a typical two-hump structure spectral energy distribution (SED) \citep{Abdo2010, Fan2016}. 
There are unique and iconic observational properties, rapid and large amplitude variability, high and variable polarization, strong and variable $\gamma$-ray emissions and apparent superluminal motion of blazars have been observed and investigated in literature \citep{Wills1992, Urry1995, Fan2002, Villata2006, Fan2014, Xiao2015, Gupta2016, Xiao2019, Xiao2020, Abdollahi2020, Fan2021}. 
These extreme observational properties are mainly due to the presence of a relativistic jet \citep{Blandford1979}, which points to the observer in a small viewing angle ($\phi$) and raises a Doppler beaming effect \citep{Ghisellini1993, Fan2013, Pei2016, Xiao2020}.
The Doppler factor ($\delta=[\Gamma(1-\beta {\rm cos}\phi)]^{-1}$, where $\Gamma$ is the bulk Lorentz factor and $\beta$ is the jet speed in units of the speed of light, $c$) is a key parameter in jets since it determines how much of flux is boosted and of timescale is compressed in the observer frame.
However, $\delta$ can only be determined indirectly because both $\beta$ and $\phi$ are unobservable quantities.
There are several indicators of the beaming effect that have been introduced in the literature, e.g. core-dominance parameter \citep{Pei2016, Pei2020} and apparent superluminal motion of knots \citep{Zhang2008, Xiao2019, Xiao2020}.

The study of knots has been performed by VLBI telescopes for decades.
These knots, especially the superluminal knots, manifest as localized intensity enhancements in radio images of jets.
The enhanced flux refers to a high brightness temperature
\begin{equation}
T = \frac{2c}{\pi k} \frac{S}{\theta_{\rm eq}^{2}\nu^{2}},
\end{equation}
where $S$ is the flux at frequency $\nu$, $\theta$ is the angular diameter of a source \citep{Readhead1994}.
The $T$ is been observed easily exceed the brightness temperature ($T_{\rm eq} \simeq 5 \times 10^{10}\, {\rm K}$) of jet equipartition, which the power of radiating particles and the power of magnetic field is comparable. 
While, the brightness temperature is also limited by the inverse Compton cooling result in a $T \leq 1.0\times 10^{12}\, {\rm K}$ \citep{Kellermann1969}.
Different authors give discrepant results of the maximum brightness temperature in the rest frame, e.g., \citet{Readhead1994} gave $3 \times 10^{11}\, {\rm K}$ and \citet{Liodakis2018} gave a similar result of $2.78 \times10^{11}\, {\rm K}$.
The early VLBI observation of brightness temperatures seems to agree with the theoretical brightness temperature limit due to inverse Compton cooling.
However, \citet{Kellermann2003} suggested that the agreement was only fortuitous and is a natural consequence of the size of the Earth and the limited range of flux density observed.


In addition to the study of its brightness temperature, the study of knot dynamics is also intriguing.
The first superluminal motion was observed on blazar 3C 279 \citep{Cohen1971}, even its explanation was proposed 5 years ago by \citet{Rees1966}.
\citet{Cohen2007} illustrated the distribution of observed apparent velocity and luminosity is consistent with relativistic beaming models, which has been widely accepted by the following researchers. 
$\beta_{\rm app}$, the apparent velocity of knots of blazar jets, is usually employed to estimate the Doppler factor and the viewing angle \citep{Xiao2019, Xiao2020} because the superluminal motion and the Doppler boosting are caused by the same geometric effect of jet orientation.
\citet{Jorstad2001a} and \citet{Kellermann2004} reported that AGN with $\gamma$-ray emission have somewhat higher observed apparent velocity based on EGRET data, and \citet{Xiao2019} found the same result with \textit{Fermi} data.
\citet{Kellermann2007} suggested there are no low luminosity sources with fast motions, but the high luminosity sources show a wide range of apparent speeds, this is consistent with the following researchers' work (e.g., \citealp{Lister2009, Piner2012, Xiao2019}).
The connection between blazar knots and the $\gamma$-ray emission has been addressed two decades ago by \citet{Jorstad2001b} based on the VLBA and EGRET data.
This idea has been further confirmed in recent works \citep{Jorstad2016, Jorstad2017, Weaver2022} by the VLBV-BU-BLAZAR Program\footnote{http://www.bu.edu/blazars/BEAM-ME.html}, which is a Boston University (BU) $\gamma$-ray blazar monitoring program at 43 GHz.

The MOJAVE\footnote{http://www.physics.purdue.edu/MOJAVE/} (Monitoring of Jets in AGN with VLBA Experiments) program is a long-term program to monitor radio brightness and polarization variations in jets associated with active galaxies visible in the northern sky.
They have published features of motions from blazar jets and reported apparent motions in many blazars \citep{Lister2013, Lister2018, Lister2019, Lister2021}.


After the launch of \textit{Fermi} large area gamma-ray space telescope (\textit{Fermi}-LAT), the study of blazars comes to its era of prosperity.
The $\gamma$-ray detection can be used to investigate the mechanism of the high energetic $\gamma$-ray origin. 
There are five generations of source catalogues that have been released, of which the latest one is the fourth \textit{Fermi}-LAT source catalogue (4FGL\_DR2, \citealp{Abdollahi2020}), in which 3511 blazars have been associated (4LAC\_DR2, \citealp{Ajello2020}).
In previous works by \citep{Xiao2019, Xiao2020}, who studied the differences between superluminal blazars with \textit{Fermi} $\gamma$-ray detection (FDS) and those without \textit{Fermi} $\gamma$-ray detection (non-FDS) by a combined use of MOJAVE and \textit{Fermi}-LAT data.
They found that the FDS shows a larger apparent velocity than the non-FDS, indicating a strong beaming effect, and suggested that the apparent velocity is an efficient indicator of the beaming effect \citep{Xiao2019, Xiao2020}.
\citet{Lister2019} found a strong correlation between apparent jet velocity and synchrotron peak frequency, with the highest jet velocities being found only in AGNs with low $\nu_{\rm p}$ values, moreover, \citet{Lister2021} analyzed the parsec-scale jet kinematics of 447 bright radio-loud AGN and suggested the \textit{Fermi}-LAT $\gamma$-ray associated AGN in MOJAVE sample tend to have more variable position angles (PAs) than the non-\textit{Fermi} AGNs.
Generally, the knots are attributed to the presence of shocks in the collimated plasma outflow \citep{Blandford1979}.
However, the correlation between knots' apparent motion and the blazar central engine is barely discussed in previous works, and the nature of how knots form and how they are accelerated still need to be explored.
In this work, we aim to investigate the connection between the apparent motion and the jet and accretion disk property and to explore if there are more candidates to be found in the future via VLBI telescopes.

This paper is organised as follows
In section 2 we present our sample and describe the data;
Our analysis and results will be presented in Section 3;
The discussions and further investigation of our results will be presented in Section 4;
We summarise our main findings in Section 5.
The cosmological parameters $H_{\rm 0} = 73 \ {\rm km \cdot s^{-1} \cdot Mpc^{-1}}$, $\Omega_{\rm m} = 0.3$ and $\Omega_{\rm \Lambda} = 0.7$ are adopted through this paper.

\section{Data Acquisition}
For the purpose of studying the property of VLBI detected blazars and the correlation between apparent motion and jet and accretion disk. 
We collect a sample of blazars with available proper motion ($\mu$) or apparent velocity ($\beta_{\rm app}$) from literature \citep{Vermeulen1994, Jorstad2005, Britzen2008, Piner2007, Piner2018, Lister2019} (see Col. (4), (5), and (6) of Tab. \ref{vfb}).
Two rules for data collecting:
(1) we use the maximum apparent velocity $\beta_{\rm app}^{\rm max}$ for the source with more than one VLBI detected component;
(2) we use the MOJAVE \citep{Lister2019} source as the base sample and omit the same source that comes from other literature because the MOJAVE program is a continuous source monitoring program and provides the latest kinematic measurements with the world's highest resolution telescope VLBA.

Then, we cross-match these sources with 4LAC\_DR2 and collect spectrum and variability information (see Col. (11), (12), (13), and (14) of Tab. \ref{vfb})).
In total, we compile a sample of 407 \textit{Fermi} blazars with available $\mu^{\rm max}$, in which 372 blazars with available $\beta_{\rm app}^{\rm max}$ from literature or calculable $\beta_{\rm app}^{\rm max}$ through
\begin{equation}
\beta_{\rm app} = {\frac{\mu}{H_{\rm 0}} {\int_1^{1+z} {\frac{1}{\sqrt{ {\Omega_{\rm M}}x^3+1-{\Omega_{\rm M}}}}}dx}}
\end{equation}
\citep{Vermeulen1994, Zhang2008, Xiao2019}.
Moreover, we also collect radio flux density ($S_{\rm \nu}$) from literature (see Col. (7), (8), and (9) of Tab. \ref{vfb})) and 5 GHz core dominance parameter (${\rm log}\,R$, Col. (10) of Tab. \ref{vfb}) from \citet{Pei2020}. 
The information of black hole mass ($M_{\rm BH}$, Col. (15) of Tab. \ref{vfb}) and accretion disk luminosity (${\rm log}\,L_{\rm Disk}$, see Col. (16) of Tab. \ref{vfb}) are collected from literature, and the intensity of the inverse Compton peak (${\rm log}\,F_{\rm IC}$, see Col. (17) of Tab. \ref{vfb}) is collected from \citet{Paliya2021}.
For the sake of convenience, these VLBI detected \textit{Fermi} blazars are denoted as VFBs in this work.

\begin{table*}
\setlength{\tabcolsep}{3pt}
\small
\caption{The sample of \textit{Fermi} blazars}
\label{vfb}
\resizebox{\textwidth}{!}{
\begin{threeparttable}
\begin{tabular}{lccccccccccccccccc}
\hline
\hline
4FGL name   &   
B1950 name  &   
z  &
ID  &   
$\mu^{\rm max}$ &
Ref &
$\nu$ &
$S^{\rm tot}$ &
Ref &
${\rm log}\, R$ &
Flux1000    &
$\alpha_{\rm ph}$   &
Class &
$VI$ &
${\rm log} \, (M_{\rm BH}/M_{\rm \odot})$ &
${\rm log} \, L_{\rm Disk}$ &
${\rm log}\, F_{\rm IC}$ &
Ref \\
    &
    &  
    &
    &   
$\mu$as/yr  &
    &
GHz &
mJy &
    &
    &
${\rm ph \cdot cm^{-2} \cdot s^{-1}}$   &
    &
    &
    &
    &
${\rm erg \cdot s^{-1}}$    &
${\rm erg \cdot cm^{-2} \cdot s^{-1}}$ \\
(1) &
(2) &
(3) &
(4) &
(5) &
(6) &
(7) &
(8) &
(9) &
(10) &
(11) &
(12) &
(13) &
(14) &
(15) &
(16) &
(17) &
(18) \\
\hline
J0005.9+3824	&	0003+380	&	0.229	&	2	&	$	317	\pm	25	$	&	L19	&	15	&	600.2	&	LI8	&	1.13	&	4.25E-10	&	2.67	&	F	&	26.77	&						&						&		&		\\
J0006.3-0620	&	0003-066	&	0.347	&	8a	&	$	330.4	\pm	9.7	$	&	L19	&	15	&	2351.9	&	LI8	&	0.26	&	1.40E-10	&	2.13	&	B	&	20.41	&	$	8.93	\pm	0.4	$	&	$	44.52	\pm	0.15	$	&	-12.04	&	P21	\\
J0009.1+0628	&	0006+061	&	1.563	&	1	&	$	134	\pm	114	$	&	L19	&	15	&	185.4	&	LI8	&		&	4.77E-10	&	2.10	&	B	&	13.05	&						&						&		&		\\
J0014.1+1910	&	0011+189	&	0.477	&	2	&	$	159	\pm	16	$	&	L19	&	15	&	133.1	&	LI8	&	0	&	2.25E-10	&	2.28	&	B	&	39.85	&	$	7.47	\pm	0.26	$	&	$	44.32	\pm	0.13	$	&	-11.65	&	P21	\\
J0014.8+6118	&	0012+610	&		&	1	&	$	13	\pm	10	$	&	L19	&		&		&		&		&	7.91E-10	&	2.05	&	U	&	6.15	&						&						&		&		\\
\hline

\multicolumn{18}{l}{Only five items are listed, this table is available in its entirety in machine-readable forms.}\\
\multicolumn{18}{l}{$^*$column (1)4FGL name;
column (2) B1959 Name;
column (3) redshift;
column (4) feature identification number;
column (5) angular proper motion in microarcseconds per year;}\\
\multicolumn{18}{l}{column (6) reference of proper motion;
column (7) frequency;
column (8) radio flux density at frequency column (7);
column (9) reference of radio flux density;}\\
\multicolumn{18}{l}{column (10) core dominance parameter at 5 GHz;
column (11) the integral photon flux from 1 to 100 GeV;
column (12) the photon spectral index;
column (13) classification,}\\
\multicolumn{18}{l}{`F' denotes the FSRQ, `B' denotes the BL Lac, and `U' denotes the blazars of uncertain type (BCU);
column (14) the variability index;
column (15) mass of central black hole;}\\
\multicolumn{18}{l}{column (16) luminosity of accretion disk;
column (17) flux of IC peak;
column (18) reference of columns (15), (16) and (17).}\\

\multicolumn{18}{l}{C99: \citet{Cao1999},
C12: \citet{Chai2012}
L18: \citet{Lister2018}, 
L19: \citet{Lister2019}, 
P21: \citet{Paliya2021}, 
P14: \citet{Piner2014}, }\\
\multicolumn{18}{l}{
P18: \citet{Piner2018}, 
Sh12: \citet{Shaw2012},
T96: \citet{Taylor1996}, 
X91: \citet{Xie1991},
Z12: \citet{Zhang2012}.}\\

\end{tabular}
\end{threeparttable}
}
\end{table*}

\section{Results}
\subsection{The correlation between maximum apparent velocity and the luminosity, photon index, and variability index of \textit{Fermi} GeV $\gamma$-ray emission}

Assuming the $\gamma$-ray photons follow a power law function and is expressed as
\begin{equation}
{\frac{dN}{dE}} = N_{\rm 0} E^{-\alpha_{\rm ph}},
\end{equation}
where $\alpha_{\rm{ph}}$ is the photon spectral index, and $N_{\rm 0}$ can be expressed as
$N_{\rm 0} = N_{(E_{\rm L}\sim E_{\rm U})}({\frac{1}{E_{\rm L}}-\frac{1}{E_{\rm U}}}),$ 
if $\alpha_{\rm ph}=2$, otherwise
$N_{\rm 0} = \frac{N_{(E_{\rm L}\sim E_{\rm U})}(1-\alpha_{\rm ph})}{(E_{\rm U}^{1-\alpha_{\rm ph}}-E_{\rm L}^{1-\alpha_{\rm ph}})},$
where $N_{(E_{\rm L}\sim E_{\rm U})}$ is the integral photons in units of ${\rm photons \cdot cm^{-2}\cdot s^{-1}}$ in the energy range of $E_{\rm L}$ - $E_{\rm U}$, where $E_{\rm L}$ and $E_{\rm U}$ correspond to 1 GeV and 100 GeV respectively.
The integral flux, $F$, in units of ${\rm GeV \cdot cm^{-2}\cdot s^{-1}}$, can be expressed in the form \citep{Fan2013, Xiao2015}
\begin{equation}
F = N_{(E_{\rm L}\sim E_{\rm U})}{\frac{E_{\rm U}-E_{\rm L}}{E_{\rm U} \times E_{\rm L}}}\ln \frac{E_{\rm U}}{E_{\rm L}}
\end{equation}
for $\alpha_{\rm ph}=2$, otherwise
\begin{equation}
F = N_{(E_{\rm L}\sim E_{\rm U})}{\frac{1-\alpha_{\rm ph}}{2-\alpha_{\rm ph}}}{\frac{(E_{\rm U}^{2-\alpha_{\rm ph}}-E_{\rm L}^{2-\alpha_{\rm ph}})}{(E_{\rm U}^{1-\alpha_{\rm ph}}-E_{\rm L}^{1-\alpha_{\rm ph}})}}.
\end{equation}
The $\gamma$-ray luminosity is calculated by
\begin{equation}
L_{\rm \gamma} = 4\pi d_{\rm L}^2(1+z)^{(\alpha_{\rm ph}-2)}F,
\end{equation}
where $d_{\rm L} = \frac{c}{H_{\rm 0}}\int^{1+z}_{1}\frac{1}{\sqrt{\Omega_{\rm m}x^{3}+1-\Omega_{\rm m}}}dx$ is a luminosity distance \citep{Komatsu2011} and $(1+z)^{(\alpha_{\rm ph}-2)}$ stands for a $K$-correction.

\begin{figure}
\centering
\includegraphics[scale=0.85]{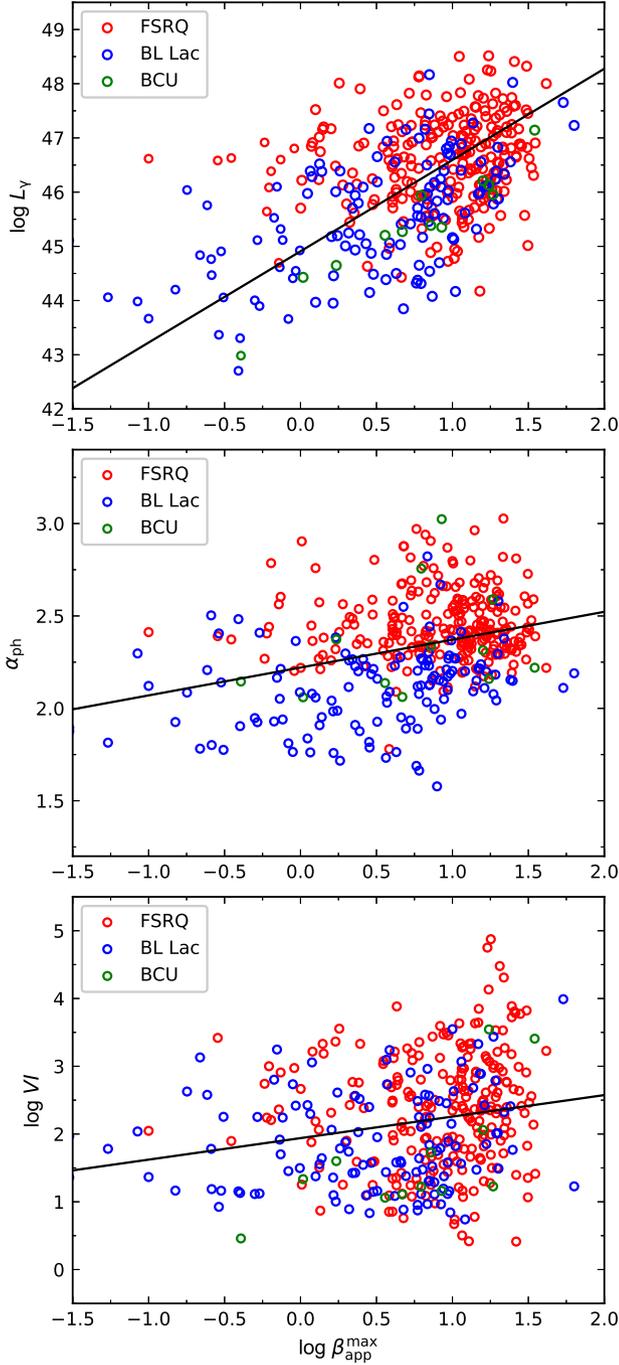}
\caption{The correlations of ${\rm log}\,L_{\rm \gamma}$, $\alpha_{\rm ph}$ and ${\rm log}\, VI$ against ${\rm log}\,\beta_{\rm app}^{\rm max}$ for the VFBs.
The FSRQs in red, the BL Lacs in blue, and the BCUs in green.}
\label{Fig_corr_3}
\end{figure}

We calculate $\gamma$-ray luminosity for the VFBs and study its correlation with apparent velocity.
The upper panel of Figure \ref{Fig_corr_3} shows the correlation between ${\rm log}\, L_{\rm \gamma}$ and ${\rm log}\, \beta_{\rm app}^{\rm max}$, and gives a linear regression
$${\rm log}\, L_{\rm \gamma}= (1.68\pm0.08){\rm log}\, \beta_{\rm app}^{\rm max} + (44.91\pm0.06),$$
and we obtained a correlation coefficient $r=0.27$ and chance probability $p=7.1 \times 10^{-8}$ through Pearson analysis (and the Pearson analysis is used to obtain $r$ and $p$ throughout this paper) when the redshift effect is removed.

The middle panel of Fig. \ref{Fig_corr_3} shows the correlation between $\alpha_{\rm ph}$ and ${\rm log}\, \beta_{\rm app}^{\rm max}$, and gives a linear regression
$$\alpha_{\rm ph}= (0.15\pm0.02){\rm log}\, \beta_{\rm app}^{\rm max} + (2.22\pm0.02),$$
with a $r=0.33$ and $p=3.1 \times 10^{-11}$.

The variability index ($VI$) is, a parameter that indicates the level of variability, introduced in the works of \textit{Fermi Collaboration}, e.g. \citet{Nolan2012} used a simple likelihood ratio to test the variability for the two-year data of \textit{Fermi} observation and \citet{Acero2015} applied the same method to the four-year data.
However, the {\textit Fermi Collaboration} added a second term (see in Eq. \ref{VI_for}) corrects (in the Gaussian limit) for the different weights between the full analysis and that in 1 yr intervals, the average flux from the light curve $F_{\rm av}$ can differ somewhat from the flux in the total analysis $F_{\rm glob}$ in the 4FGL, expressed as:
\begin{equation}
VI = 2 \sum_{i} {\rm log} \left[ \frac{\mathcal{L}_{i}(F_{i})}{\mathcal{L}_{i}(F_{\rm glob})} \right] - {\rm max(\chi^{2}(F_{\rm glob}) - \chi^{2}(F_{\rm av}), 0)},
\label{VI_for}
\end{equation}
\begin{equation}
\chi^{2}(F) = \sum_{i} \frac{(F_{i}-F)^{2}}{\sigma_{i}^{2}},
\end{equation}
where $F_{i}$ are the individual flux values, $\mathcal{L}_{i}(F)$ is the likelihood in the interval $i$ assuming flux $F$, and $\sigma_{i}$ are the errors on $F_{i}$, $F_{\rm av}$ is the average flux over one-year interval light curve and $F_{\rm glob}$ is the flux obtained from a global fit.
A source would be considered to show probable variability if $VI > 18.48$, corresponding to 99\% confidence in a $\chi^{2}$ distribution, at GeV band \citep{Abdollahi2020}.
In the bottom panel of Figure \ref{Fig_corr_3} shows the correlation between ${\rm log}\, VI$ and ${\rm log}\, \beta_{\rm app}^{\rm max}$, and gives a linear regression
$${\rm log}\, VI = (0.32\pm0.07){\rm log}\, \beta_{\rm app}^{\rm max} + (1.94\pm0.07),$$
with a $r=0.21$ and $p=2.7 \times 10^{-5}$.

\subsection{The correlation between maximum apparent velocity and black hole mass, accretion disk luminosity, and the normalized accretion luminosity}
\begin{figure}
\centering
\includegraphics[scale=0.85]{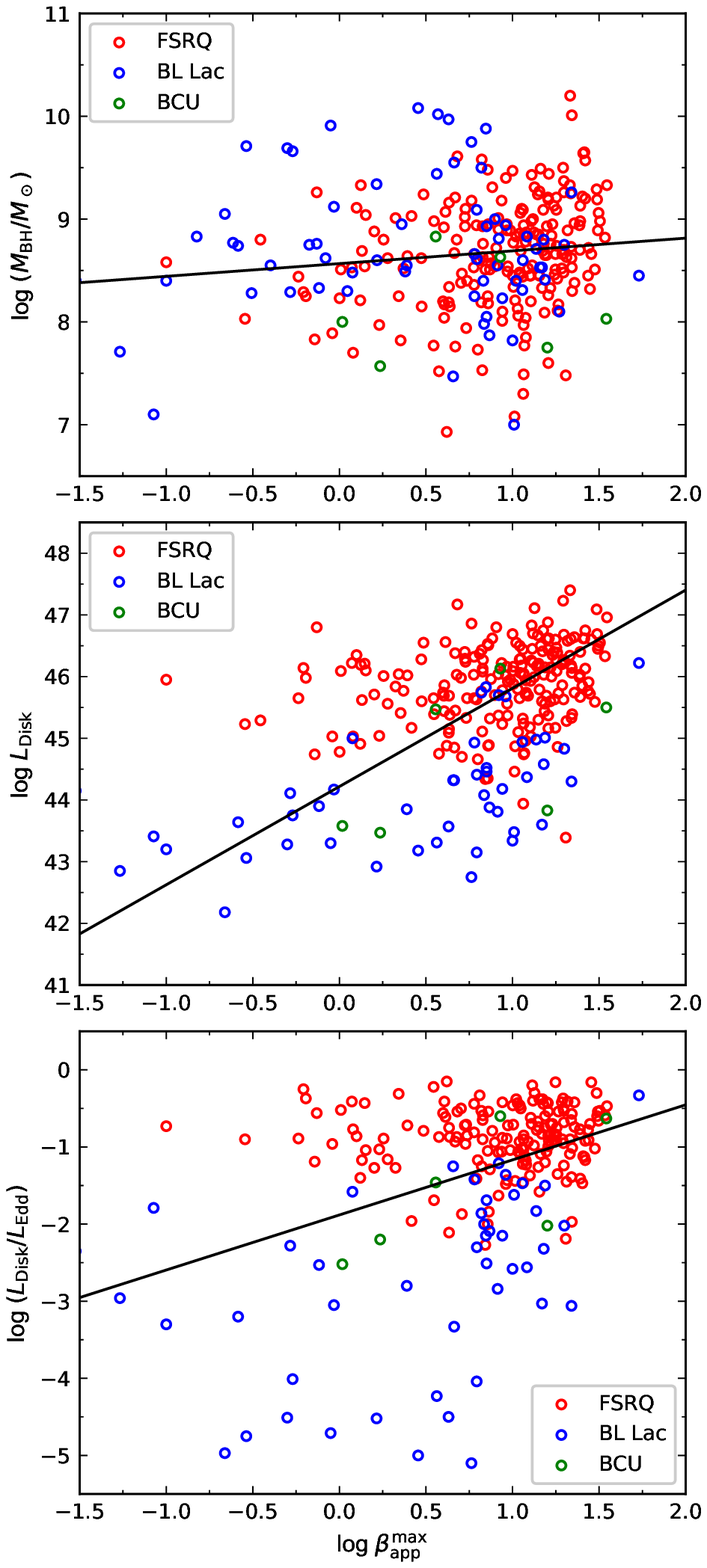}
\caption{The correlations of ${\rm log} \, (M_{\rm BH}/M_{\rm \odot})$, ${\rm log}\, L_{\rm Disk}$, and ${\rm log}\,(L_{\rm Disk}/L_{\rm Edd})$ against ${\rm log}\,\beta_{\rm app}^{\rm max}$ for the VFBs.
The FSRQs in red, the BL Lacs in blue, and the BCUs in green.}
\label{Fig_corr_4}
\end{figure}

The black hole mass is estimated in several ways.
The typical tactic is to use the virial theorem and estimate the central black hole mass through
\begin{equation}
M_{\rm BH} \approx rv^{2}/G,
\end{equation}
where $v$ is the velocity dispersion of matter at distance $r$.
The velocity dispersion can be derived from spectroscopic study and expressed as $v=fv_{\rm FWHM}$, where $v_{\rm FWHM}$ is the FWHM of the broad emission line and $f$ is, an geometry and kinematics dependent factor, on the order of unity \citep{Peterson1999, Peterson2000, McLure2001, Vestergaard2002}.
The distance $r$ can be well determined by the reverberation mapping, and so that the `Size-Luminosity Relation' \citep{Kaspi2000}.
Thus, the virial $M_{\rm BH}$ can be calculated using the following equation \citep{Shen2011, Paliya2021}:
\begin{equation}
{\rm log}\, \left( \frac{M_{\rm BH}}{M_{\odot}} \right) = a +b \, {\rm log}\, \left( \frac{\lambda L_{\rm \lambda}}{10^{44} \, {\rm erg \cdot s^{-1}}} \right) + 2 \, {\rm log}\, \left( \frac{\rm FWHM}{\rm km \cdot s^{-1}} \right),
\end{equation}
where $\lambda L_{\rm \lambda}$ is the continuum luminosity at given wavelength, e.g., 5100 $\mathring{\rm A}$ for H${\rm \beta}$, 3000 $\mathring{\rm A}$ for M\lowercase{g}\,{\sevensize II} and 1350 $\mathring{\rm A}$ for C\,{\sevensize IV}.
The coefficients $a$ and $b$ can be taken from \citet{McLure2004} and \citet{Vestergaard2006}.
Similarly, a stellar velocity dispersion is used to estimate $M_{\rm BH}$ \citep{Gultekin2009}.
Moreover, the host galaxy bulge luminosity is also suggested to be the estimator of $M_{\rm BH}$.
\citet{Paliya2021} utilized these three methods and calculated $M_{\rm BH}$ for a sample of \textit{Fermi} blazars, in which 269 sources in our sample are included.
The upper panel of Fig. \ref{Fig_corr_4} shows the correlation between ${\rm log}\, (M_{\rm BH}/M_{\odot})$ and ${\rm log}\,\beta^{\rm max}_{\rm app}$, gives a linear regression
$${\rm log}\, (M_{\rm BH}/M_{\odot}) = (0.12 \pm 0.06) {\rm log}\,\beta^{\rm max}_{\rm app} + (8.57 \pm 0.06),$$
with $r=0.12$ and $p=0.05$.

The accretion disk luminosity ($L_{\rm Disk}$) is usually derived either from the SED modelling of the blue bump or from the broad emission-line luminosity ($L_{\rm BLR} = 0.1\, L_{\rm Disk}$), where the 0.1 is the broad emission-line region (BLR) covering factor.
The $L_{\rm BLR}$ can be calculated by using the following equation
\begin{equation}
L_{\rm BLR} = \sum_{i} L_{i} \cdot \frac{\langle L_{\rm BLR, \ rel} \rangle}{\sum_{i} L_{i,\ \rm rel}},
\end{equation}
where $\langle L_{\rm BLR,\ rel} \rangle = 556$, $L_{i}$ is observed line luminosity, and $L_{i,\ \rm rel}$ is relative line luminosity.
The Ly$\alpha$ as a reference flux that contributed to 100, the relative weight of H$\alpha$, H$\beta$, M\lowercase{g}\,{\sevensize II} and C\,{\sevensize IV} lines to 77, 22, 34, and 63, the total broad emission-line flux is then fixed as 556 \citep{Celotti1997, Francis1991, Paliya2021, Xiao2022a}.
\citet{Paliya2021} utilized the methods and calculated $L_{\rm Disk}$ or $3\sigma$ upper limits for a sample \textit{Fermi} blazars, in which 254 sources of our sample are included.

In the middle panel of Fig. \ref{Fig_corr_4}, the plot shows a positive correlation between ${\rm log}\, L_{\rm Disk}$ and ${\rm log}\, \beta_{\rm app}^{\rm max}$ gives a linear regression
$${\rm log}\, L_{\rm Disk}= (1.59\pm 0.10){\rm log}\, \beta_{\rm app}^{\rm max} + (44.22\pm0.08),$$
with a correlation coefficient $r=0.32$ and chance probability $p=1.1 \times 10^{-7}$ when the redshift effect is removed.

The Eddington luminosity $L_{\rm Edd}$ is a function of Eddington accretion rate (${\dot M}_{\rm Edd}$) and is expressed as $L_{\rm Edd} = {\dot M}_{\rm Edd} c^{2} = 1.26 \times 10^{38} (M/M_{\rm \odot}) \ {\rm erg \cdot s^{-1}}$.
We calculate $L_{\rm Edd}$ for the sources with black hole mass in our sample, analysis the correlation between ${\rm log}\, (L_{\rm Disk}/L_{\rm Edd})$ and ${\rm log}\,\beta^{\rm max}_{\rm app}$ as shown in the lower panel of Fig. \ref{Fig_corr_4}, in which a linear regression gives
$${\rm log}\, (L_{\rm Disk}/L_{\rm Edd}) = (0.71 \pm 0.11) {\rm log}\,\beta^{\rm max}_{\rm app} - (1.88 \pm 0.11),$$
with $r=0.40$ and $p=4.8 \times 10^{10}$.

\subsection{The distributions of VFBs and the rest of \textit{Fermi} blazars}
\begin{figure}
\centering
\includegraphics[scale=0.85]{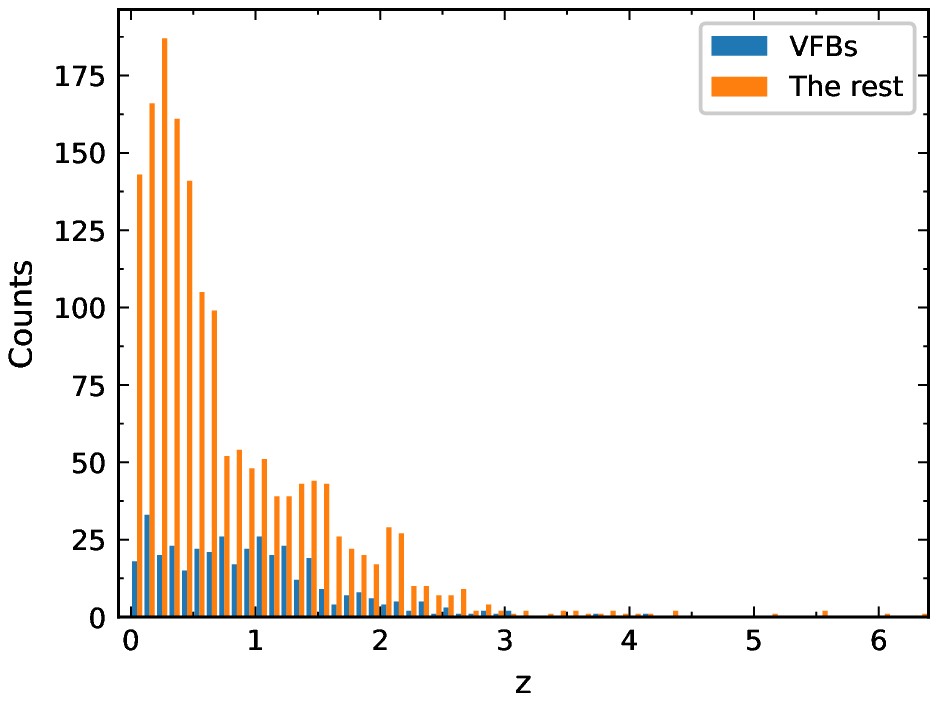}
\caption{The redshift $z$ distribution for the VFBs (the blue bar) and the rest of \textit{Fermi} blazars (the orange bar).}
\label{Fig_hist_z}
\end{figure}

The 4FGL\_R2 contains 3511 blazars, in which 2009 sources with available redshift in literature or from NED.
The redshift ranges from 0.00351 (4FGL J1248.3+5820/PG 1246+586) to 4.162481 (4FGL J0929.3+5014/GB6 J0929+5013) with an average value of $\langle z^{\rm VFB} \rangle = 0.95 \pm 0.68$ for the VFBs, while redshift ranges from 0.00001 (4FGL J1113.8+1528/2MASS J11135586+1528058) to 6.3952 (4FGL J1233.7-0144/NVSS J123341-014426) with an average value of $\langle z^{\rm R} \rangle = 0.79 \pm 0.76$ for the rest of \textit{Fermi} blazars, as shown in Fig. \ref{Fig_hist_z}.

\begin{figure}
\centering
\includegraphics[scale=0.85]{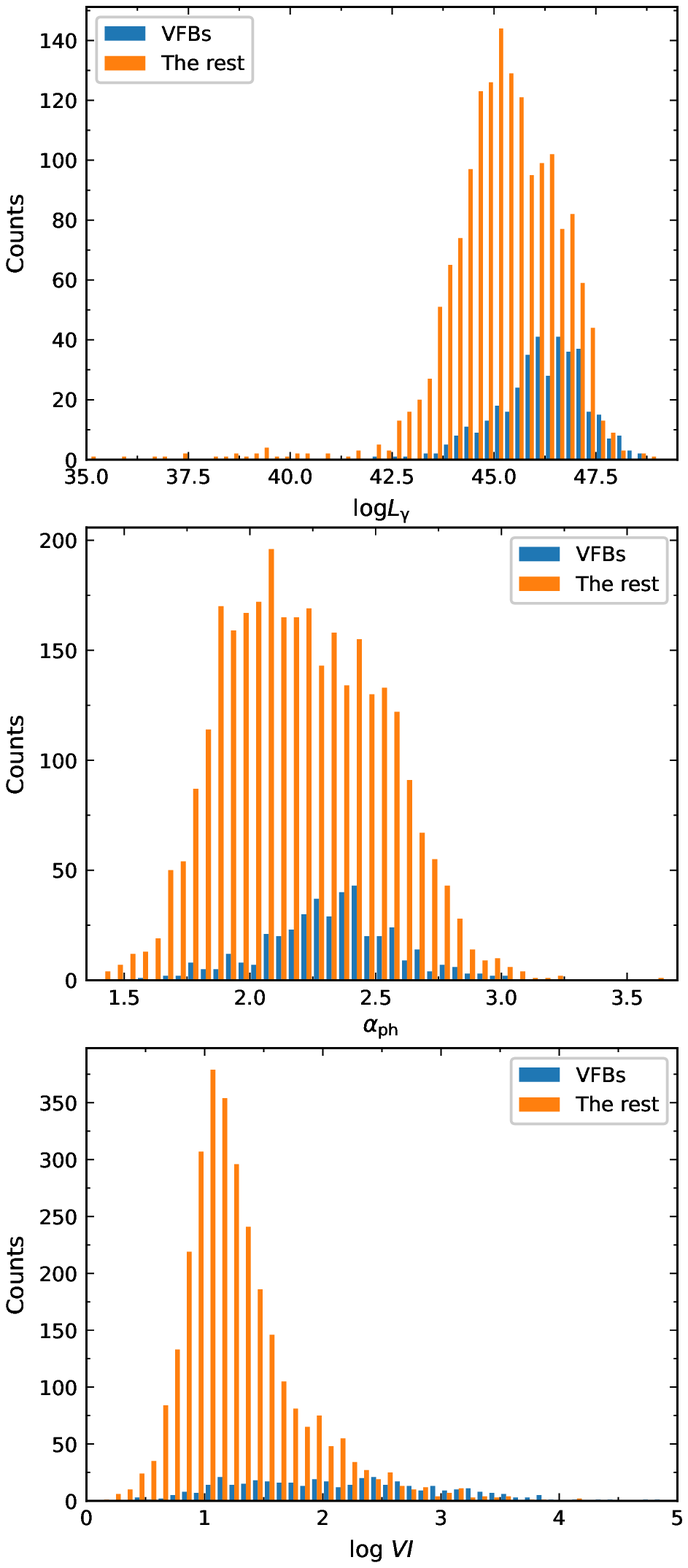}
\caption{The ${\rm log}\,L_{\rm \gamma}$, $\alpha_{\rm ph}$ and ${\rm log}\, VI$ distribution for the VFBs (the blue bar) and the rest of \textit{Fermi} blazars (the orange bar).}
\label{Fig_hist}
\end{figure}

We calculate the $\gamma$-ray luminosity for both the VFBs and the rest of \textit{Fermi} blazars, and illustrate it in the upper panel of Fig. \ref{Fig_hist}.
The luminosity of the VFBs ranges from 42.04 to 48.51 with an average value $\langle {\rm log}L_{\rm \gamma}^{\rm VFB} \rangle= 46.20 \pm 1.07$, and ranges from 35.22 to 48.76 with an average luminosity of the rest blazars is $\langle {\rm log}L_{\rm \gamma}^{\rm R} \rangle = 45.24 \pm 1.42$.
Anderson-Darling (A-D) test is applied to test if ${\rm log}L_{\rm \gamma}^{\rm VFB}$ and ${\rm log}L_{\rm \gamma}^{\rm R}$ are form the same distribution.
The A-D test gives a statistic 111.3 greater than the critical statistic 6.5 for significance level 0.1\%, the null hypothesis that the two samples come from the same distribution.
Thus we can state the $\gamma$-ray luminosity distributions for the VFBs is different from that of the rest \textit{Fermi} blazars, and the VFBs have a larger average $\gamma$-ray luminosity than the rest.

The photon index ($\alpha_{\rm ph}$) ranges from 1.58 to 3.03 with an average value $\langle \alpha_{\rm ph}^{\rm VFB}\rangle = 2.33 \pm 0.25$ for the VFBs;
and $\alpha_{\rm ph}$ ranges from 1.40 to 3.63 with an average value $\langle \alpha_{\rm ph}^{\rm R}\rangle = 2.21 \pm 0.30$ for the rest \textit{Fermi} blazars, as shown in the middle panel of Figure \ref{Fig_hist}.
An A-D test is applied to the distributions of $\alpha_{\rm ph}^{\rm VFB}$ and $\alpha_{\rm ph}^{\rm R}$, the A-D test gives a statistic 47.0 greater than the critical statistic 6.5 for significance level 0.1\%, suggesting $\alpha_{\rm ph}^{\rm VFB}$ and $\alpha_{\rm ph}^{\rm R}$ are from different distributions, and the VFBs have larger photon index than the rest of \textit{Fermi} blazars.


In our sample, ${\rm log}\,VI$ ranges from 0.41 to 4.88 with an average value $\langle {\rm log}\,VI^{\rm VFB} \rangle = 2.14 \pm 0.84$ for the VFBs;
${\rm log}\,VI$ ranges from 0.16 to 4.13 with an average value $\langle {\rm log}\,VI^{\rm R} \rangle = 1.30 \pm 0.50$ for the rest of \textit{Fermi} blazars, seen in the bottom panel of Fig. \ref{Fig_hist}.
An A-D test is applied to the distributions of ${\rm log}\,VI^{\rm VFB}$ and ${\rm log}\,VI^{\rm R}$, the A-D test gives a statistic 327.9 greater than the critical statistic 6.5 for significance level 0.1\%, suggesting ${\rm log}\,VI^{\rm VFB}$ and ${\rm log}\,VI^{\rm R}$ are from different distributions, and the VFBs have larger variability index than the rest of \textit{Fermi} blazars.


\section{Discussion}

\subsection{The apparent motion and the central engine of blazars}
\begin{figure}
\centering
\includegraphics[scale=0.85]{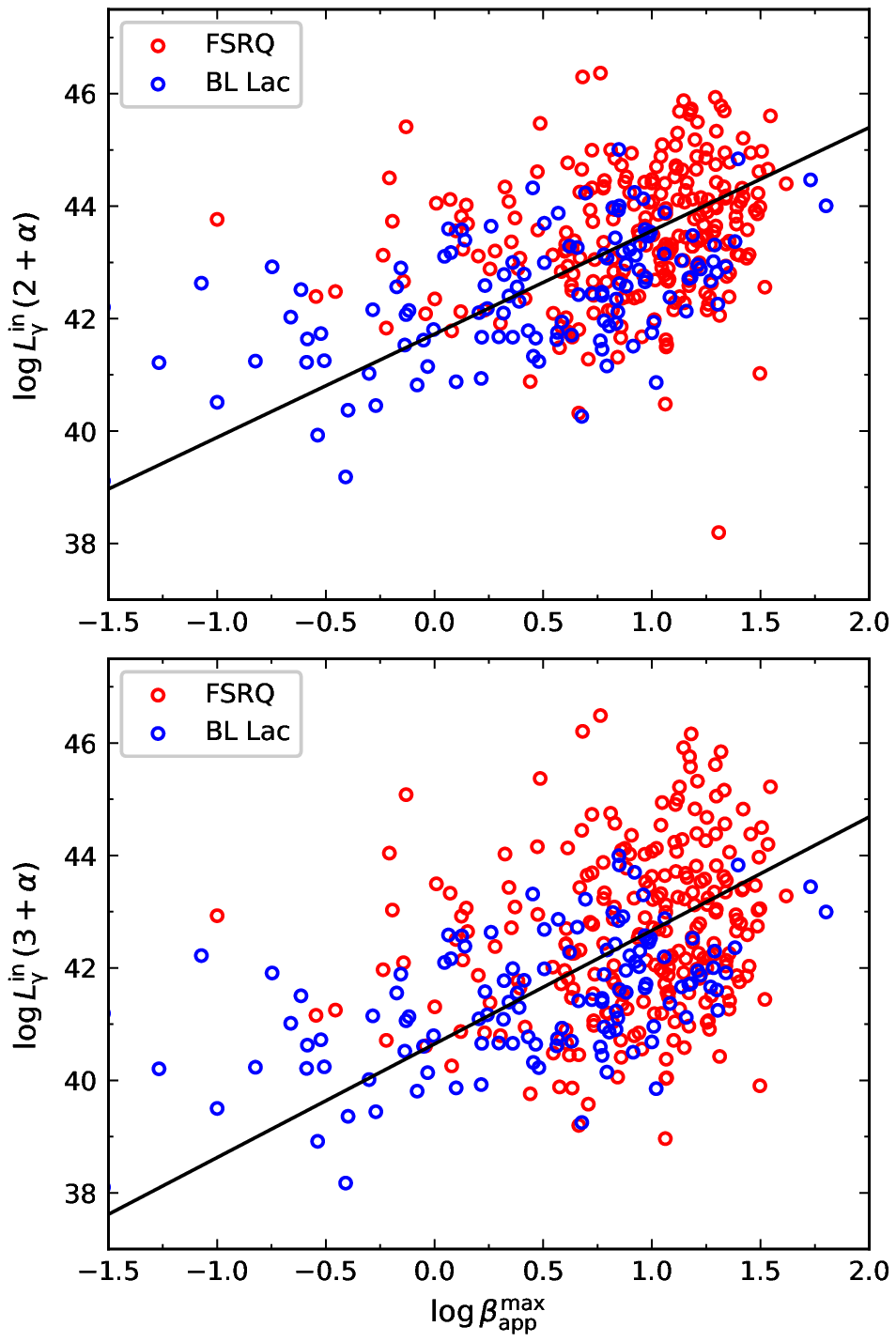}
\caption{The correlations of intrinsic ${\rm log}\,L_{\rm \gamma}^{\rm in}$ against ${\rm log}\,\beta_{\rm app}^{\rm max}$ for the VFBs.
The FSRQs in red and the BL Lacs in blue.}
\label{Fig_corr_in}
\end{figure}

We have studied the correlation between maximum apparent velocity and the GeV $\gamma$-ray for the VFBs in our sample.
It is shown that a positive correlation between ${\rm log}\,L_{\rm \gamma}$ and ${\rm log}\, \beta_{\rm app}^{\rm max}$.
Before discussing the implications of this correlation, there is a caveat to note, concerning the important Doppler beaming effect.
Blazar $\gamma$-ray emission has been proved to be boosted due to a beaming effect \citep{Xiao2015, Pei2016, Fan2016, Xiao2019, Xiao2020, Zhang2020, Paliya2021}.
We extract the $L_{\rm \gamma}$ regarding the Doppler beaming effect according to
\begin{equation}
L_{\rm \gamma}^{\rm in} = L_{\rm \gamma}^{\rm ob}/\delta^{q+1},
\label{eq_d}
\end{equation}
where $\delta$ is the Doppler factor that we calculated with the approach suggested in \citet{Zhang2020}, $q = 2 + \alpha$ for a continuous jet emission (or $q=3+\alpha$ for a discrete jet emission), and $\alpha$ ($=\alpha_{\rm ph}-1$) is a spectral index $f_{\nu} \propto \nu^{-\alpha}$.
A correlation of the intrinsic $\gamma$-ray luminosity ${\rm log}\,L_{\rm \gamma}^{\rm in}$ against ${\rm log}\,\beta_{\rm app}^{\rm max}$ is shown in Fig. \ref{Fig_corr_in}, the linear regression gives
$${\rm log}\, L_{\rm \gamma}^{\rm in}= (1.84\pm0.10){\rm log}\, \beta_{\rm app}^{\rm max} + (41.72\pm0.08),$$
with a $r=0.26$ and $p=5.8 \times 10^{-7}$ for the case of $q=2+\alpha$ when the redshift effect is removed; and 
$${\rm log}\, L_{\rm \gamma}^{\rm in}= (2.02\pm0.13){\rm log}\, \beta_{\rm app}^{\rm max} + (40.65\pm0.10),$$
with a $r=0.23$ and $p=1.0 \times 10^{-5}$ for the case of $q=3+\alpha$ when the redshift effect is removed.
Obviously, the positive correlations (${\rm log}\,\beta_{\rm app}^{\rm max} \, vs \, {\rm log}\,L_{\rm \gamma}$ and ${\rm log}\,\beta_{\rm app}^{\rm max} \, vs \, {\rm log}\,L_{\rm \gamma}^{\rm in}$) indicating that the motion of knots may correlate with the jet central engine.

In general, the entire jet power ($P_{\rm jet}$) is attributed to the radiation power ($P_{\rm rad}$) and the kinetic power ($P_{\rm kin}$), which is responsible for jet nonthermal radiation and its propagation, respectively.
It is possible to estimate $P_{\rm rad}$ and $P_{\rm kin}$ through several methods, among of which the broadband SED fitting based on the simultaneous data gives both the kinetic and the radiation power \citep{Ghisellini2014, Tan2020}.
\citet{Ghisellini2010} and \citet{Ghisellini2014} carried out the radiation power
\begin{equation}
P_{\rm rad} = 2f\frac{\Gamma^{2} L_{\rm jet}^{\rm bol}}{\delta^{4}},
\label{eq_P_rad}
\end{equation}
where the factor 2 counts for two-sided jets, $f$ equals 16/5 for the case of SSC process (mostly BL Lacs),  $f=4/3$ and replace $\delta^{4}$ with $\delta^{4}(\delta/\Gamma)^2$ for the case of EC process (mostly FSRQs).
We can calculate $P_{\rm rad}$ with two assumptions.
Firstly, $\Gamma$ is believed to be equal to $\delta$ for blazars because of the small viewing angle \citep{Ghisellini2010, Ghisellini2014, Xiong2014, Zhang2020, Xiao2022a}.
Secondly, the $L_{\rm \gamma}$ has been involved in calculating $\delta$ through the method in \citet{Zhang2020}, thus for the purpose of avoiding degeneration between $L_{\rm \gamma}$ and $\delta$, seen in Eq. \ref{eq_P_rad}, the $L_{\rm bol}^{\rm jet}$ is represented by the inverse Compton peak luminosity $L_{\rm IC}$ rather than $L_{\rm \gamma}$ as used in previous works \citep{Ghisellini2010, Ghisellini2014, Xiong2014, Zhang2020}.
The peak luminosity of inverse Compton is given by $L_{\rm IC} = 4\pi d_{\rm L}^{2}\, F_{\rm IC}$, where $F_{\rm IC}$, the inverse Compton peak flux, is obtained from \citet{Paliya2021}.

\begin{figure}
\centering
\includegraphics[scale=0.85]{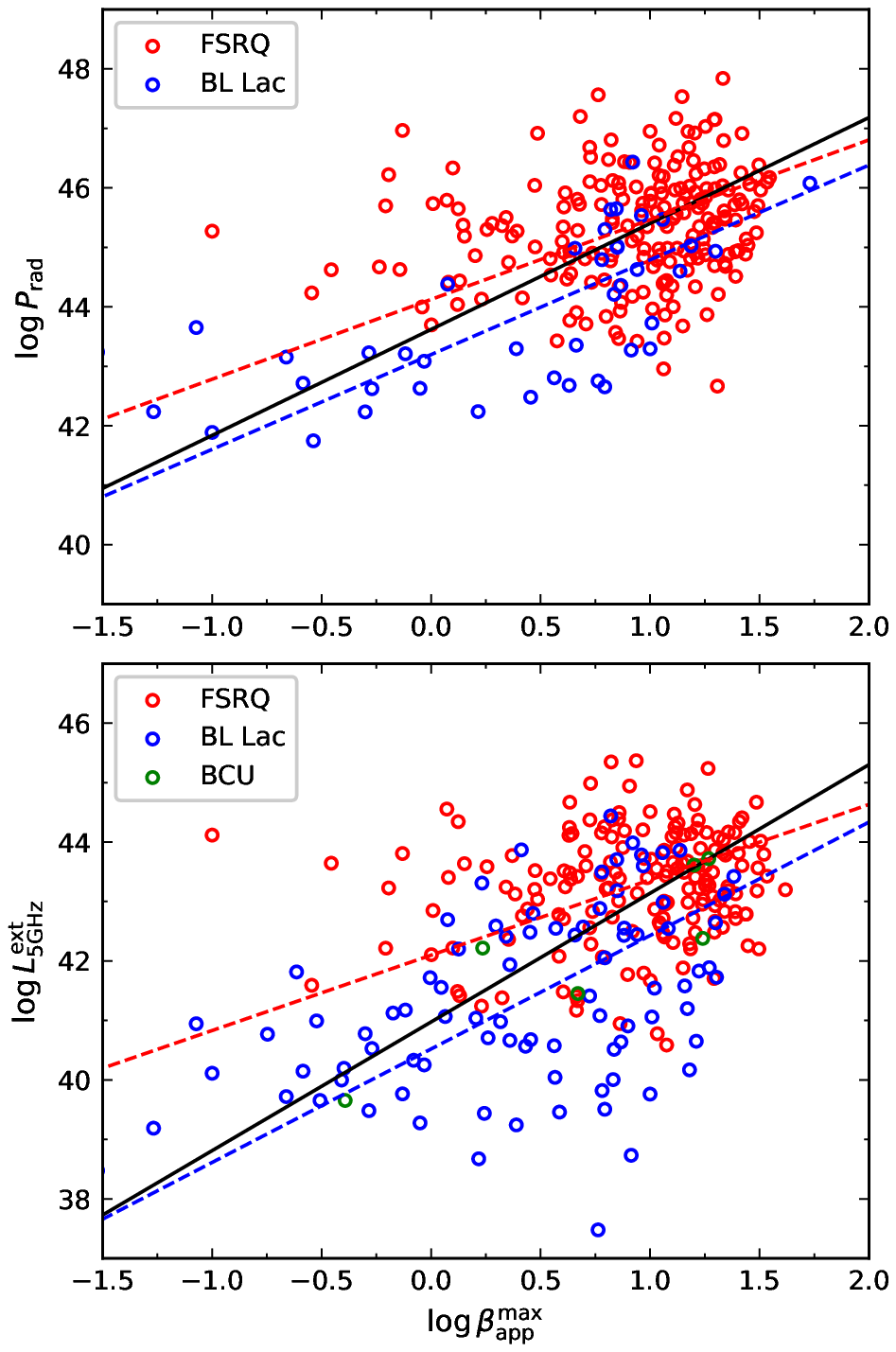}
\caption{The correlations of ${\rm log}\,P_{\rm rad}$ and ${\rm log}\,L_{\rm 5GHz}^{\rm ext}$ against ${\rm log}\,\beta_{\rm app}^{\rm max}$ for the VFBs.
The FSRQs in red, the BL Lacs in blue, and the BCUs in green.}
\label{Fig_corr_jet}
\end{figure}

\citet{Cavagnolo2010} suggested that the jet kinetic power is able to inflate the X-ray cavities or bubbles in different systems including giant elliptical galaxies and cD galaxies (Type cD galaxy, a subtype of type-D giant elliptical galaxy), and proposed to evaluate the kinetic power $P_{\rm kin}=P_{\rm cav}$.
However, this method is only limited to a small number of sources at present.
It is known that the luminosity of extended region of radio jet, which is believed to be less Doppler boosted, is related to jet kinetic power \citep{Rawlings1991, Willott1999, Cavagnolo2010, Meyer2011}, $P_{\rm rad} = \eta \, (L_{\rm 5GHz}^{\rm ext})^{\kappa}$.
Through The factor $\kappa$ and $\eta$ are given in discrepancy in literature due to the different sizes and source types of sample \citep{Cavagnolo2010, Meyer2011}, the ${\rm log}\,L^{\rm ext}_{\rm 5GHz}$ scales with the ${\rm log}\,P_{\rm rad}$ in the logarithmic space.
We collect the total radio flux density from literature (\citealp{Taylor1996} at 5 GHz, \citealp{Piner2014} at 8.4 GHz, and \citealp{Lister2018} at 15 GHz) and we convert the data at other frequencies to 5 GHz by assuming that
\begin{equation}
S_{\rm 5GHz}^{\rm core} = S_{\rm \nu}^{\rm core} \,\, {\rm and} \,\, S_{\rm 5GHz}^{\rm ext} = S_{\rm \nu}^{\rm ext} \left( \frac{\nu}{\rm 5 \, GHz} \right)^{\alpha_{\rm ext}},
\end{equation}
where the total radio flux is the sum of the flux of core and the flux of the extended region, $S^{\rm tot} = S^{\rm core} + S^{\rm ext}$, the $\alpha_{\rm ext} = 0.75$ and $\alpha_{\rm core} = 0$ \citep{Fan2011, Pei2016, Pei2019, Pei2020}.
Together with the radio core dominance parameter at 5 GHz
\begin{equation}
R = \left( \frac{S^{\rm core}}{S^{\rm ext}} \right) (1+z)^{\alpha_{\rm core}-\alpha_{\rm ext}},
\end{equation}
that we collect from \citet{Pei2020}, we obtain $S_{\rm 5GHz}^{\rm ext}$ and calculate ${\rm log}\,L^{\rm ext}_{\rm 5GHz}$.
The correlations of ${\rm log}\,P_{\rm rad}$ and ${\rm log}\,L_{\rm 5GHz}^{\rm ext}$ against ${\rm log}\,\beta_{\rm app}^{\rm max}$ are illustrated in Fig. \ref{Fig_corr_jet} and linear regression results are listed in Tab. \ref{lin_cor_jet}.
Positive correlations of ${\rm log}\,P_{\rm rad}\, vs \, {\rm log}\,\beta_{\rm app}^{\rm max}$ and ${\rm log}\,L_{\rm 5GHz}^{\rm ext}\, vs \, {\rm log}\,\beta_{\rm app}^{\rm max}$ are found for blazars.
The positive correlation of ${\rm log}\,P_{\rm rad}\, vs \, {\rm log}\,\beta_{\rm app}^{\rm max}$ holds for both FSRQs and BL Lacs when we consider them independently, while the positive correlation of ${\rm log}\,L_{\rm 5GHz}^{\rm ext}\, vs \, {\rm log}\,\beta_{\rm app}^{\rm max}$ only hold for BL Lacs.
It is found that the motion of jet knots is significantly correlated with jet radiation power for both FSRQs and BL Lacs, however, the motion of jet knots is correlated with the kinetic power only for BL Lacs.

\begin{table*}
\centering
\setlength{\tabcolsep}{16pt}
\renewcommand{\arraystretch}{1.3}
\caption{The correlation of ${\rm log}\,P_{\rm rad}$ and ${\rm log}\,L_{\rm 5GHz}^{\rm ext}$ against ${\rm log}\,\beta_{\rm app}^{\rm max}$}
\label{lin_cor_jet}
\begin{tabular}{lcccccc}
\hline
\hline
Type & class & $(a \pm \Delta a)$ & $(b \pm \Delta b)$ & $N$ & $r$ & $p$\\
\hline
& All & 1.78 $\pm$ 0.12 & 43.62 $\pm$ 0.10 & 249 & 0.29 & $5.0 \times 10^{-6}$  \\
${\rm log}\,P_{\rm rad}\, vs \,{\rm log}\,\beta_{\rm app}^{\rm max}$& FSRQs & 1.34 $\pm$ 0.12 & 44.12 $\pm$ 0.11 & 207 & 0.19 & $6.9 \times 10^{-3}$ \\
& BL Lacs & 1.59 $\pm$ 0.19 & 43.20 $\pm$ 0.07 & 42 & 0.34 & 0.03 \\
\hline
& All & 2.16 $\pm$ 0.12 & 40.97 $\pm$ 0.09 & 270 & 0.27 & $1.0 \times 10^{-5}$  \\
${\rm log}\,L_{\rm 5GHz}^{\rm ext}\, vs \,{\rm log}\,\beta_{\rm app}^{\rm max}$& FSRQs & 1.27 $\pm$ 0.15 & 42.10 $\pm$ 0.14 & 176 & 0.12 & 0.11 \\
& BL Lacs & 1.91 $\pm$ 0.16 & 40.52 $\pm$ 0.07 & 88 & 0.25 & 0.02 \\
\hline
\end{tabular}
\end{table*}

Moreover, we have investigated the correlation between apparent motion and black hole mass, luminosity of accretion disk, and normalized disk luminosity, seen in Fig. \ref{Fig_corr_4}.
Our results suggest the ${\rm log}\,\beta_{\rm app}^{\rm max}$ is not correlated with ${\rm log}\, (M_{\rm BH}/M_{\odot})$ because of a $p=0.05$.
While the correlations between ${\rm log}\,\beta_{\rm app}^{\rm max}$, and ${\rm log}\,L_{\rm Disk}$, and ${\rm log}\,(L_{\rm Disk}/L_{\rm Edd})$ are moderately significant, the positive correlations suggest that stronger accretion disk tend to have faster apparent velocities of knots.
The results indicate that the central black hole mass may not be a key parameter of a jet to generate knots and to accelerate them, while the power of accretion disk and the accretion rate is likely to be critical to the jet.
Previous studies has established the blazar knot propagation disturbance to be connected with $\gamma$-ray flares, as well as the jet activities \citep{Jorstad2001b, Marscher2010, Jorstad2013, Tanaka2015, Jorstad2016, Jorstad2017}.
In the majority of these cases, $\gamma$-ray outbursts coincide with the passage of superluminal knots through the radio core \citep{Jorstad2016}.
In this scheme, we may find a way of the accretion disk connecting the relativistic jet.
We suggest a sudden increase of accretion rate of a powerful accretion disk would generate and accelerate knots, and the propagation of knot disturbance derive jet activities and show outbursts.

\subsection{The criteria of selecting VFB candidates from the \textit{Fermi} blazars}
Blazars show various observational properties, for instance, the high $\gamma$-ray luminosity.
In the GeV $\gamma$-ray band, the spectrum is a superposition of different radiation mechanisms, e.g., the scattering of synchroton photons (SSC), of the seed photons from accretion disk \citep{Dermer1993}, of the ultraviolet seed photons from broad line region (BLR, \citealp{Sikora1994}), of the seed photons from dusty torus (DT, \citealp{Blazejowski2000, Arbeiter2002, Sokolov2005}) by the relativistic electrons via inverse Compton process.
Therefore, sources should show difference in the spectrum if their $\gamma$-ray emission consists of different mechanisms.
\citet{Ackermann2015} found that FSRQs show greater $\gamma$-ray luminosity $L_{\rm \gamma}$ and larger photon index $\alpha_{\rm ph}$ than BL Lacs by analyzing the third catalog of AGNs detected by the \textit{Fermi}-LAT (3LAC).
Variation is one of the characters of blazars, \citet{Yang2022} suggests that FSRQs have larger ${\rm log} VI$ than the BL Lacs.
Our results suggest that ${\rm log}\,\beta_{\rm app}^{\rm max}$ is positively correlated with both $\alpha_{\rm ph}$ and ${\rm log}\, VI$, seen in the middle and bottom panels of Fig. \ref{Fig_corr_3}, indicate blazar have faster apparent motion tend to have softer $\gamma$-ray spectrum and higher magnitude of variability.

For the facts that we have mentioned above, in the present work, we investigate the differences between the VLBI detected \textit{Fermi} blazars and the rest of \textit{Fermi} blazars through these three parameters.
Our results suggest that the VFBs have larger ${\rm log}\, L_{\rm \gamma}$, $\alpha_{\rm ph}$ and ${\rm log}\, VI$ than the rest of \textit{Fermi} blazars.

Based on the complete MOJAVE 1.5 Jy sample of AGNs, \cite{Lister2015} found that 23\% of these AGNs are not detected by \textit{Fermi} because of an instrumental selection effect and partly due to their lower Doppler boosting factors.
\citet{Xiao2019} suggested that those blazars not detected by \textit{Fermi}-LAT should be \textit{Fermi} blazar candidates if they have large apparent velocity, we also predicted some sources to be \textit{Fermi} blazar candidates and this prediction is proved to be true \citep{Xiao2020} when the 4FGL released.
On one side, the blazars with apparent motion detected through VLBI technique are likely to be \textit{Fermi} $\gamma$-ray source candidates.
On the other side, whether there are more blazars in the \textit{Fermi} catalogue are VLBI candidates?
There are 3437 blazars that have been identified in 4FGL\_DR2, out of which 407 sources (taking 11.8\%) are associated with VLBI detection and most of them are detected by MOJAVE program. 
The quantity of VFBs is obviously underestimated because the MOJAVE program is only dedicated to monitoring the sources in the northern sky (and the southern sky source near the equator), while \textit{Fermi}-LAT gives all-sky observation of $\gamma$-ray emitters.


For the purpose of finding more VFBs, we explore criteria of selecting VFB candidates from 4FGL based on the ${\rm log}\,L_{\rm \gamma}$, $\alpha_{\rm ph}$, and ${\rm log}\,VI$.

\begin{figure*}
\centering
\includegraphics[scale=1.0]{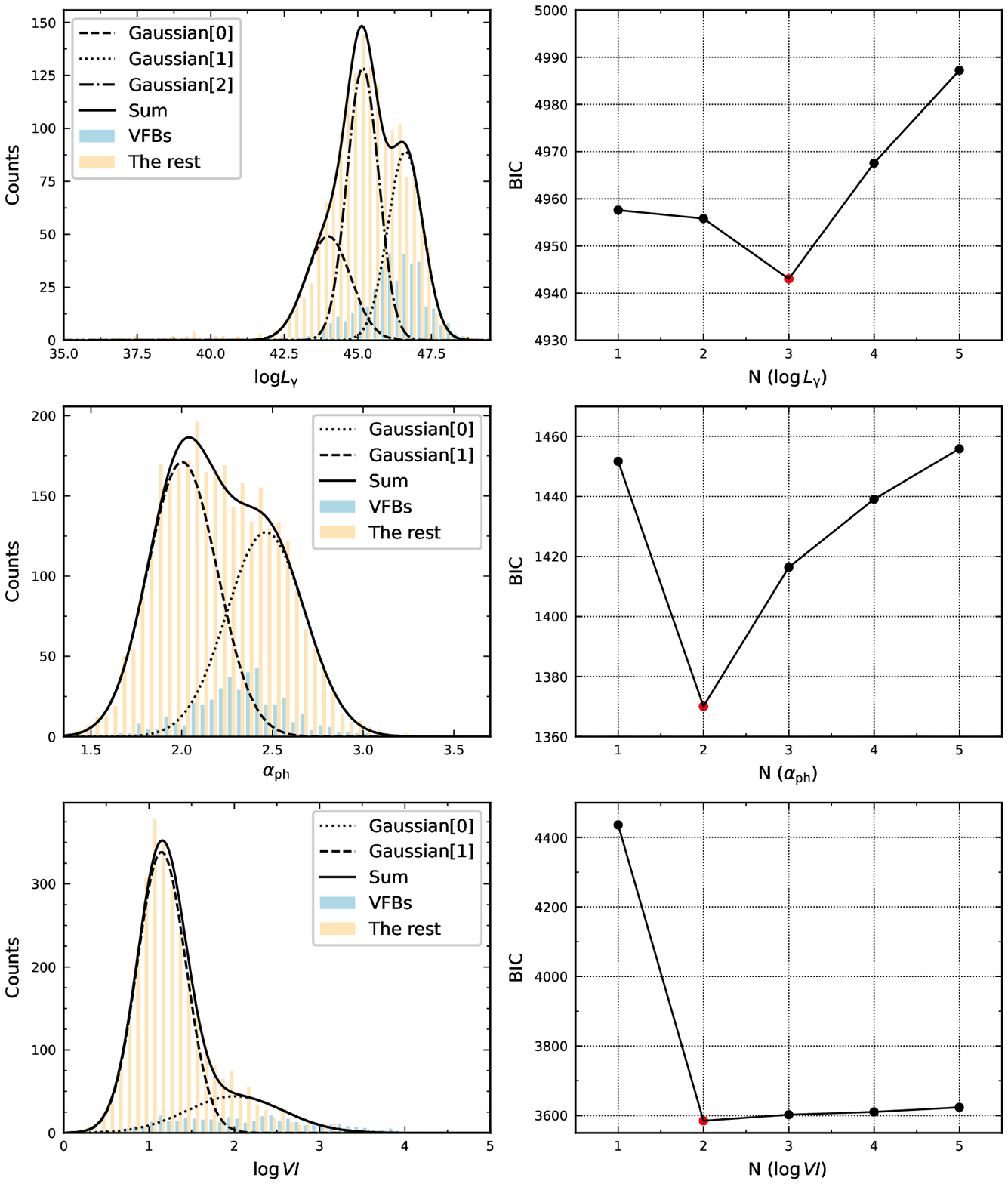}
\caption{The distributions of $\gamma$-ray luminosity (${\rm log}\,L_{\rm \gamma}$), photon index ($\alpha_{\rm ph}$) and variability index (${\rm log}\,VI$) that fitted with \textit{GMM}. 
The blue bar stands for the VFBs, the orange bar stands for the rest of \textit{Fermi} blazars.}
\label{Fig_gmm}
\end{figure*}

We apply \textit{Gaussian mixture models} (GMM), which is a probabilistic model that assumes all the data points are generated from a mixture of a finite number of Gaussian distributions with unknown parameters in the package of \textit{sklearn}, which is a software machine learning library for the \textit{Python} \citep{Fraley2002}, to decompose ${\rm log}L_{\rm \gamma}$, $\alpha_{\rm ph}$ and ${\rm log}\,VI$ distributions as shown in the left panel of Fig. \ref{Fig_gmm}.
The proportions of each Gaussian component of the total distribution (the hidden variables) and the mean and standard deviation of each Gaussian component are obtained through an \textit{Expectation-Maximization} (EM) algorithm, which is an iterative algorithm. 
In the `E' step, the initial value of the parameters or the values of the last iteration is used to calculate the posterior probability of hidden variables.
In the `M' step, the likelihood function is maximized to obtain the mean and standard deviation.
The `E' and `M' steps are iterated until all the estimated parameters are converged.
We apply stratified sampling with a scale of 95\% to run the whole GMM process 100 times in order to increase the generalization ability of our model.
\textit{Bayesian Information Criterion} (BIC) is employed to qualify the GMM model and quantify the number of \textit{Gaussian} components.
BIC value is a criterion for model selection among a finite set of models, models with lower BIC are generally preferred.
In this work, we employ two criteria, which are the lowest BIC value and $\Delta {\rm BIC}> 10$ \citep{Kass1995}, to select the decomposition model.

The BIC result suggests a three \textit{Gaussian} components, $N=3$ gives the lowest BIC=4943.0 and $\Delta {\rm BIC}_{\rm 32} = 12.8$, for the distribution of ${\rm log}\,L_{\rm \gamma}$ of the rest of \textit{Fermi} blazars.
The distribution gives clustering probability of 0.25 for component \textit{Gaussian}[0], 0.32 for component \textit{Gaussian}[1] and 0.43 for component \textit{Gaussian}[2], see in the top-right panel of Fig. \ref{Fig_gmm}.

The BIC result suggests a two \textit{Gaussian} components, $N=2$ gives the lowest BIC=1370.1 and $\Delta {\rm BIC}_{\rm 32} = 35.$, for the distribution of $\alpha_{\rm ph}$ of the rest of \textit{Fermi} blazars.
The distribution gives clustering probability of 0.47 for component \textit{Gaussian}[0] and 0.53 for component \textit{Gaussian}[1], see in the middle-right panel of Fig. \ref{Fig_gmm}.

A two Gaussian components model, $N=2$ gives the lowest BIC=3585.0 and $\Delta {\rm BIC}_{\rm 32} = 17.4$, which gives clustering probability of 0.21 for component \textit{Gaussian}[0] and 0.79 for component \textit{Gaussian}[1], is found for ${\rm log}\,VI$ for the rest of \textit{Fermi} blazars as shown in the bottom-right panel of Fig. \ref{Fig_gmm}.

It is shown that the VFBs are mainly enveloped by \textit{Gaussian}[1] and \textit{Gaussian}[2] for ${\rm log}\,L_{\rm \gamma}$, are almost enveloped by \textit{Gaussian}[0] for $\alpha_{\rm ph}$ and enveloped by \textit{Gaussian}[0] for ${\rm log}\,VI$.
We propose that the dotted-curve envelope the `VFB-like' sources, the dash-dotted curve concludes the `transition' sources and the dashed-curve contains the `non-VFB-like' sources and suggest that the conjunction point of the dotted-curve and dashed-curve as the boundary of selecting VFB candidates.
Our calculation gives the conjunction point coordinates, ${\rm log}\,L_{\rm \gamma} = 45.40 \pm 0.22$, $\alpha_{\rm ph} = 2.24 \pm 0.01$ and ${\rm log}\,VI = 1.71 \pm 0.07$.
There are 228 sources predicted as VFB candidates, whose luminosity, photon index and variability index are greater than their boundary, and they are listed in Tab. \ref{candi}.

We notice that there are 79 VFBs (taking 19.4\%) with ${\rm log}\,L_{\rm \gamma} < 45.40$, 144 VFBs (taking 35.4\%) with $\alpha_{\rm ph} < 2.24$, and 142 VFBs (taking 34.9\%) with ${\rm log}\,VI < 1.71$ corresponds to $VI=51.29$. 
Thus, one should be aware that there are a significant number of VFBs with ${\rm log}\,L_{\rm \gamma}$ and/or $\alpha_{\rm ph}$ and/or ${\rm log}\,VI$ less than the boundary value that we proposed, our method and criteria is proposed for selecting more promising VFB candidates.

Among our 228 VFB candidates, there are 88 (38.6\%) sources in the northern sky and 140 (61.4\%) sources in the southern hemisphere.
This result should help the VLBI equipment to find more blazars with jet `knot' structure and monitor their kinetics in the future, especially for MOJAVE.

\begin{table*}
\centering
\caption{The superluminal \textit{Fermi} blazar candidates}
\label{candi}
\begin{threeparttable}
\begin{tabular}{lcccccc}
\hline
4FGL Name   & Class & z & Flux1000 & ${\rm log}\,L_{\rm \gamma}$ & $\alpha_{\rm ph}$ & $VI$ \\
 & & & ${\rm photon \cdot cm^{-2} \cdot s^{-1}}$ & ${\rm erg \cdot cm^{-2} \cdot s^{-1}}$ & \\
\hline
4FGL J0001.5+2113	&	fsrq	&	1.106	&	1.36E-09	&	46.74	&	2.66	&	1564.42	\\
4FGL J0004.4-4737	&	fsrq	&	0.88	&	4.36E-10	&	45.99	&	2.37	&	139.12	\\
4FGL J0010.6-3025	&	fsrq	&	1.19	&	3.49E-10	&	46.24	&	2.46	&	91.59	\\
4FGL J0011.4+0057	&	fsrq	&	1.492	&	4.29E-10	&	46.60	&	2.33	&	71.98	\\
4FGL J0016.2-0016	&	fsrq	&	1.57631	&	4.17E-10	&	46.68	&	2.74	&	82.14	\\
\hline

\multicolumn{7}{l}{$^{*}$column (1) gives the 4FGL name;
column (2) gives the classification;
column (3) gives the redshift;}\\
\multicolumn{7}{l}{column (4) gives the integral photon flux from 1 to 100 GeV;
column (5) gives the $\gamma$-ray luminosity;}\\
\multicolumn{7}{l}{column (6) gives the photon spectral index;
column (7) gives the variability index.}\\
\multicolumn{7}{l}{Only five items are listed, this table is available in its entirety in machine-readable forms.}\\

\end{tabular}
\end{threeparttable}
\end{table*}

\section{Conclusion}
In this work, we made use of a sample of VLBI detected \textit{Fermi} blazars to study the blazar jet and accretion disk property.
We found that 
(1) a robust correlation between the maximum apparent velocity (${\rm log}\, \beta^{\rm max}_{\rm app}$) and $\gamma$-ray luminosity (${\rm log}\, L_{\rm \gamma}$), and this correlation is intrinsic because the correlation still exists when we removed the Doppler beaming effect;
(2) following the ${\rm log}\, L_{\rm \gamma}\, {\rm vs}\, {\rm log}\, \beta^{\rm max}_{\rm app}$, we notice knot apparent motion is related to the jet power and a larger ${\rm log}\, \beta^{\rm max}_{\rm app}$ should correspond to a more powerful jet;
we found both FSRQs and BL Lacs have ${\rm log}\, \beta^{\rm max}_{\rm app}$ correlated with the jet radiation power, however, only BL Lacs have ${\rm log}\, \beta^{\rm max}_{\rm app}$ correlated with the 5 GHz jet extended region luminosity (${\rm log}\, L_{\rm 5GHz}^{\rm ext}$), which is an indicator of jet kinetic power;
(3) ${\rm log}\, \beta^{\rm max}_{\rm app}$ shows no correlation with black hole mass (${\rm log}\,(M_{\rm BH}/M_{\odot})$), while it has a moderate correlation with accretion disk luminosity (${\rm log}\, L_{\rm Disk}$) and the normalized accretion disk luminosity (${\rm log}\, (L_{\rm Disk}/L_{\rm Edd})$).
The result suggests both the power of accretion disk and the accretion rate are critical to generate and accelerate blazar knots.
We suggest one way of connecting accretion disk and relativistic jet that the knots are caused by sudden increase of accretion rate of a powerful accretion disk and the propagation of knots disturbance would arise jet activities and show outbursts.  

In addition, we found (1)
the VFBs have differences between the rest of \textit{Fermi} blazars on $\gamma$-ray luminosity (${\rm log}\, L_{\rm \gamma}$), $\gamma$-ray photon index ($\alpha_{\rm ph}$), and variability index (${\rm log}\, VI$).
The VFBs have average values of 
$\langle {\rm log}L_{\rm \gamma}^{\rm VFB} \rangle= 46.20 \pm 1.07$, 
$\langle \alpha^{\rm VFB} \rangle= 2.33 \pm 0.25$,
$\langle {\rm log} VI^{\rm VFB} \rangle= 2.14 \pm 0.84$,
while the average values for the rest of \textit{Fermi} blazars are 
$\langle {\rm log}L_{\rm \gamma}^{\rm R} \rangle = 45.24 \pm 1.42$,
$\langle \alpha^{\rm VFB} \rangle= 2.21 \pm 0.30$,
$\langle {\rm log} VI^{\rm VFB} \rangle= 1.30 \pm 0.50$;
(2) based on the difference of ${\rm log}\, L_{\rm \gamma}$, $\alpha_{\rm ph}$ and ${\rm log}\, VI$, we generated a criteria to select VFB candidates from the rest of \textit{Fermi} blazars by using a \textit{Gaussian mixture models} (GMM) method.
We suggested a blazar has ${\rm log}\,L_{\rm \gamma} > 45.40$, $\alpha_{\rm ph} > 2.24$ and ${\rm log}\,VI > 1.71$ can be considered as VFB candidates.
According to this criteria, we managed to select 228 VFBs.

\section*{Acknowledgements}
We thank the support of the key laboratory for astrophysics of Shanghai.
H. B, Xiao acknowledges the support from National Natural Science Foundation of China (NSFC 12203034) and from Shanghai science and Technology Fund (22YF1431500);
S. H, Zhang acknowledges the support from the Natural Science Foundation of Shanghai (20ZR1473600);
J. H, Fan acknowledges the support from the NSFC (NSFC U2031201, NSFC 11733001);
Z. J, Luo acknowledges the supported by Shanghai science and Technology Fund (Grant No.20070502400).
We would like to thank the MOJAVE team for the use of their precious kinematic data.

\section*{Data Availability}
For the study presented in this work, we used the following published data:
(1)4FGL, \url{https://fermi.gsfc.nasa.gov/ssc/data/access/lat/10yr_catalog/} and 4LAC.
(2)MOJAVE Program, \url{http://www.physics.purdue.edu/astro/MOJAVE/allsources.html}, and literature in this paper.
(3)NED, \url{https://ned.ipac.caltech.edu/forms/byname.html}.


\bibliographystyle{mnras}
\bibliography{lib} 

\begin{thebibliography}{}
\makeatletter
\relax
\def\mn@urlcharsother{\let\do\@makeother \do\$\do\&\do\#\do\^\do\_\do\%\do\~}
\def\mn@doi{\begingroup\mn@urlcharsother \@ifnextchar [ {\mn@doi@}
  {\mn@doi@[]}}
\def\mn@doi@[#1]#2{\def\@tempa{#1}\ifx\@tempa\@empty \href
  {http://dx.doi.org/#2} {doi:#2}\else \href {http://dx.doi.org/#2} {#1}\fi
  \endgroup}
\def\mn@eprint#1#2{\mn@eprint@#1:#2::\@nil}
\def\mn@eprint@arXiv#1{\href {http://arxiv.org/abs/#1} {{\tt arXiv:#1}}}
\def\mn@eprint@dblp#1{\href {http://dblp.uni-trier.de/rec/bibtex/#1.xml}
  {dblp:#1}}
\def\mn@eprint@#1:#2:#3:#4\@nil{\def\@tempa {#1}\def\@tempb {#2}\def\@tempc
  {#3}\ifx \@tempc \@empty \let \@tempc \@tempb \let \@tempb \@tempa \fi \ifx
  \@tempb \@empty \def\@tempb {arXiv}\fi \@ifundefined
  {mn@eprint@\@tempb}{\@tempb:\@tempc}{\expandafter \expandafter \csname
  mn@eprint@\@tempb\endcsname \expandafter{\@tempc}}}

\bibitem[\protect\citeauthoryear{{Abdo} et~al.,}{{Abdo}
  et~al.}{2010}]{Abdo2010}
{Abdo} A.~A.,  et~al., 2010, \mn@doi [\apj] {10.1088/0004-637X/716/1/30}, \href
  {https://ui.adsabs.harvard.edu/abs/2010ApJ...716...30A} {716, 30}

\bibitem[\protect\citeauthoryear{{Abdollahi} et~al.,}{{Abdollahi}
  et~al.}{2020}]{Abdollahi2020}
{Abdollahi} S.,  et~al., 2020, \mn@doi [\apjs] {10.3847/1538-4365/ab6bcb},
  \href {https://ui.adsabs.harvard.edu/abs/2020ApJS..247...33A} {247, 33}

\bibitem[\protect\citeauthoryear{{Acero} et~al.,}{{Acero}
  et~al.}{2015}]{Acero2015}
{Acero} F.,  et~al., 2015, \mn@doi [\apjs] {10.1088/0067-0049/218/2/23}, \href
  {https://ui.adsabs.harvard.edu/abs/2015ApJS..218...23A} {218, 23}

\bibitem[\protect\citeauthoryear{{Ackermann} et~al.,}{{Ackermann}
  et~al.}{2015}]{Ackermann2015}
{Ackermann} M.,  et~al., 2015, \mn@doi [\apj] {10.1088/0004-637X/810/1/14},
  \href {https://ui.adsabs.harvard.edu/abs/2015ApJ...810...14A} {810, 14}

\bibitem[\protect\citeauthoryear{{Ajello} et~al.,}{{Ajello}
  et~al.}{2020}]{Ajello2020}
{Ajello} M.,  et~al., 2020, \mn@doi [\apj] {10.3847/1538-4357/ab791e}, \href
  {https://ui.adsabs.harvard.edu/abs/2020ApJ...892..105A} {892, 105}

\bibitem[\protect\citeauthoryear{{Arbeiter}, {Pohl}  \&
  {Schlickeiser}}{{Arbeiter} et~al.}{2002}]{Arbeiter2002}
{Arbeiter} C.,  {Pohl} M.,   {Schlickeiser} R.,  2002, \mn@doi [\aap]
  {10.1051/0004-6361:20020221}, \href
  {https://ui.adsabs.harvard.edu/abs/2002A&A...386..415A} {386, 415}

\bibitem[\protect\citeauthoryear{{Blandford} \& {Koenigl}}{{Blandford} \&
  {Koenigl}}{1979}]{Blandford1979}
{Blandford} R.~D.,  {Koenigl} A.,  1979, \aplett, \href
  {https://ui.adsabs.harvard.edu/abs/1979ApL....20...15B} {20, 15}

\bibitem[\protect\citeauthoryear{{B{\l}a{\.z}ejowski}, {Sikora}, {Moderski}  \&
  {Madejski}}{{B{\l}a{\.z}ejowski} et~al.}{2000}]{Blazejowski2000}
{B{\l}a{\.z}ejowski} M.,  {Sikora} M.,  {Moderski} R.,   {Madejski} G.~M.,
  2000, \mn@doi [\apj] {10.1086/317791}, \href
  {https://ui.adsabs.harvard.edu/abs/2000ApJ...545..107B} {545, 107}

\bibitem[\protect\citeauthoryear{{Britzen} et~al.,}{{Britzen}
  et~al.}{2008}]{Britzen2008}
{Britzen} S.,  et~al., 2008, \mn@doi [\aap] {10.1051/0004-6361:20077717}, \href
  {https://ui.adsabs.harvard.edu/abs/2008A&A...484..119B} {484, 119}

\bibitem[\protect\citeauthoryear{{Cao} \& {Jiang}}{{Cao} \&
  {Jiang}}{1999}]{Cao1999}
{Cao} X.,  {Jiang} D.~R.,  1999, \mn@doi [\mnras]
  {10.1046/j.1365-8711.1999.02657.x}, \href
  {https://ui.adsabs.harvard.edu/abs/1999MNRAS.307..802C} {307, 802}

\bibitem[\protect\citeauthoryear{{Cavagnolo}, {McNamara}, {Nulsen}, {Carilli},
  {Jones}  \& {B{\^\i}rzan}}{{Cavagnolo} et~al.}{2010}]{Cavagnolo2010}
{Cavagnolo} K.~W.,  {McNamara} B.~R.,  {Nulsen} P.~E.~J.,  {Carilli} C.~L.,
  {Jones} C.,   {B{\^\i}rzan} L.,  2010, \mn@doi [\apj]
  {10.1088/0004-637X/720/2/1066}, \href
  {https://ui.adsabs.harvard.edu/abs/2010ApJ...720.1066C} {720, 1066}

\bibitem[\protect\citeauthoryear{{Celotti}, {Padovani}  \&
  {Ghisellini}}{{Celotti} et~al.}{1997}]{Celotti1997}
{Celotti} A.,  {Padovani} P.,   {Ghisellini} G.,  1997, \mn@doi [\mnras]
  {10.1093/mnras/286.2.415}, \href
  {https://ui.adsabs.harvard.edu/abs/1997MNRAS.286..415C} {286, 415}

\bibitem[\protect\citeauthoryear{{Chai}, {Cao}  \& {Gu}}{{Chai}
  et~al.}{2012}]{Chai2012}
{Chai} B.,  {Cao} X.,   {Gu} M.,  2012, \mn@doi [\apj]
  {10.1088/0004-637X/759/2/114}, \href
  {https://ui.adsabs.harvard.edu/abs/2012ApJ...759..114C} {759, 114}

\bibitem[\protect\citeauthoryear{{Cohen}, {Cannon}, {Purcell}, {Shaffer},
  {Broderick}, {Kellermann}  \& {Jauncey}}{{Cohen} et~al.}{1971}]{Cohen1971}
{Cohen} M.~H.,  {Cannon} W.,  {Purcell} G.~H.,  {Shaffer} D.~B.,  {Broderick}
  J.~J.,  {Kellermann} K.~I.,   {Jauncey} D.~L.,  1971, \mn@doi [\apj]
  {10.1086/151204}, \href
  {https://ui.adsabs.harvard.edu/abs/1971ApJ...170..207C} {170, 207}

\bibitem[\protect\citeauthoryear{{Cohen}, {Lister}, {Homan}, {Kadler},
  {Kellermann}, {Kovalev}  \& {Vermeulen}}{{Cohen} et~al.}{2007}]{Cohen2007}
{Cohen} M.~H.,  {Lister} M.~L.,  {Homan} D.~C.,  {Kadler} M.,  {Kellermann}
  K.~I.,  {Kovalev} Y.~Y.,   {Vermeulen} R.~C.,  2007, \mn@doi [\apj]
  {10.1086/511063}, \href
  {https://ui.adsabs.harvard.edu/abs/2007ApJ...658..232C} {658, 232}

\bibitem[\protect\citeauthoryear{{Dermer} \& {Schlickeiser}}{{Dermer} \&
  {Schlickeiser}}{1993}]{Dermer1993}
{Dermer} C.~D.,  {Schlickeiser} R.,  1993, \mn@doi [\apj] {10.1086/173251},
  \href {https://ui.adsabs.harvard.edu/abs/1993ApJ...416..458D} {416, 458}

\bibitem[\protect\citeauthoryear{{Fan}}{{Fan}}{2002}]{Fan2002}
{Fan} J.-H.,  2002, \mn@doi [\pasj] {10.1093/pasj/54.4.L55}, \href
  {https://ui.adsabs.harvard.edu/abs/2002PASJ...54L..55F} {54, L55}

\bibitem[\protect\citeauthoryear{{Fan}, {Yang}, {Pan}  \& {Hua}}{{Fan}
  et~al.}{2011}]{Fan2011}
{Fan} J.-H.,  {Yang} J.-H.,  {Pan} J.,   {Hua} T.-X.,  2011, \mn@doi [Research
  in Astronomy and Astrophysics] {10.1088/1674-4527/11/12/004}, \href
  {https://ui.adsabs.harvard.edu/abs/2011RAA....11.1413F} {11, 1413}

\bibitem[\protect\citeauthoryear{{Fan}, {Yang}, {Liu}  \& {Zhang}}{{Fan}
  et~al.}{2013}]{Fan2013}
{Fan} J.-H.,  {Yang} J.-H.,  {Liu} Y.,   {Zhang} J.-Y.,  2013, \mn@doi
  [Research in Astronomy and Astrophysics] {10.1088/1674-4527/13/3/002}, \href
  {https://ui.adsabs.harvard.edu/abs/2013RAA....13..259F} {13, 259}

\bibitem[\protect\citeauthoryear{{Fan}, {Bastieri}, {Yang}, {Liu}, {Hua},
  {Yuan}  \& {Wu}}{{Fan} et~al.}{2014}]{Fan2014}
{Fan} J.-H.,  {Bastieri} D.,  {Yang} J.-H.,  {Liu} Y.,  {Hua} T.-X.,  {Yuan}
  Y.-H.,   {Wu} D.-X.,  2014, \mn@doi [Research in Astronomy and Astrophysics]
  {10.1088/1674-4527/14/9/004}, \href
  {https://ui.adsabs.harvard.edu/abs/2014RAA....14.1135F} {14, 1135}

\bibitem[\protect\citeauthoryear{{Fan} et~al.,}{{Fan} et~al.}{2016}]{Fan2016}
{Fan} J.~H.,  et~al., 2016, \mn@doi [\apjs] {10.3847/0067-0049/226/2/20}, \href
  {https://ui.adsabs.harvard.edu/abs/2016ApJS..226...20F} {226, 20}

\bibitem[\protect\citeauthoryear{{Fan} et~al.,}{{Fan} et~al.}{2021}]{Fan2021}
{Fan} J.~H.,  et~al., 2021, \mn@doi [\apjs] {10.3847/1538-4365/abd32d}, \href
  {https://ui.adsabs.harvard.edu/abs/2021ApJS..253...10F} {253, 10}

\bibitem[\protect\citeauthoryear{{Fraley} \& {Raftery}}{{Fraley} \&
  {Raftery}}{2002}]{Fraley2002}
{Fraley} C.,  {Raftery} A E.,  2002, \mn@doi [Journal of the American
  Statistical Association] {10.1198/016214502760047131}, 97, 611

\bibitem[\protect\citeauthoryear{{Francis}, {Hewett}, {Foltz}, {Chaffee},
  {Weymann}  \& {Morris}}{{Francis} et~al.}{1991}]{Francis1991}
{Francis} P.~J.,  {Hewett} P.~C.,  {Foltz} C.~B.,  {Chaffee} F.~H.,  {Weymann}
  R.~J.,   {Morris} S.~L.,  1991, \mn@doi [\apj] {10.1086/170066}, \href
  {https://ui.adsabs.harvard.edu/abs/1991ApJ...373..465F} {373, 465}

\bibitem[\protect\citeauthoryear{{Ghisellini} \& {Tavecchio}}{{Ghisellini} \&
  {Tavecchio}}{2010}]{Ghisellini2010}
{Ghisellini} G.,  {Tavecchio} F.,  2010, \mn@doi [\mnras]
  {10.1111/j.1745-3933.2010.00952.x}, \href
  {https://ui.adsabs.harvard.edu/abs/2010MNRAS.409L..79G} {409, L79}

\bibitem[\protect\citeauthoryear{{Ghisellini}, {Padovani}, {Celotti}  \&
  {Maraschi}}{{Ghisellini} et~al.}{1993}]{Ghisellini1993}
{Ghisellini} G.,  {Padovani} P.,  {Celotti} A.,   {Maraschi} L.,  1993, \mn@doi
  [\apj] {10.1086/172493}, \href
  {https://ui.adsabs.harvard.edu/abs/1993ApJ...407...65G} {407, 65}

\bibitem[\protect\citeauthoryear{{Ghisellini}, {Tavecchio}, {Maraschi},
  {Celotti}  \& {Sbarrato}}{{Ghisellini} et~al.}{2014}]{Ghisellini2014}
{Ghisellini} G.,  {Tavecchio} F.,  {Maraschi} L.,  {Celotti} A.,   {Sbarrato}
  T.,  2014, \mn@doi [\nat] {10.1038/nature13856}, \href
  {https://ui.adsabs.harvard.edu/abs/2014Natur.515..376G} {515, 376}

\bibitem[\protect\citeauthoryear{{G{\"u}ltekin} et~al.,}{{G{\"u}ltekin}
  et~al.}{2009}]{Gultekin2009}
{G{\"u}ltekin} K.,  et~al., 2009, \mn@doi [\apj] {10.1088/0004-637X/698/1/198},
  \href {https://ui.adsabs.harvard.edu/abs/2009ApJ...698..198G} {698, 198}

\bibitem[\protect\citeauthoryear{{Gupta} et~al.,}{{Gupta}
  et~al.}{2016}]{Gupta2016}
{Gupta} A.~C.,  et~al., 2016, \mn@doi [\mnras] {10.1093/mnras/stw377}, \href
  {https://ui.adsabs.harvard.edu/abs/2016MNRAS.458.1127G} {458, 1127}

\bibitem[\protect\citeauthoryear{{Jorstad} \& {Marscher}}{{Jorstad} \&
  {Marscher}}{2016}]{Jorstad2016}
{Jorstad} S.,  {Marscher} A.,  2016, \mn@doi [Galaxies]
  {10.3390/galaxies4040047}, \href
  {https://ui.adsabs.harvard.edu/abs/2016Galax...4...47J} {4, 47}

\bibitem[\protect\citeauthoryear{{Jorstad}, {Marscher}, {Mattox}, {Wehrle},
  {Bloom}  \& {Yurchenko}}{{Jorstad} et~al.}{2001a}]{Jorstad2001a}
{Jorstad} S.~G.,  {Marscher} A.~P.,  {Mattox} J.~R.,  {Wehrle} A.~E.,  {Bloom}
  S.~D.,   {Yurchenko} A.~V.,  2001a, \mn@doi [\apjs] {10.1086/320858}, \href
  {https://ui.adsabs.harvard.edu/abs/2001ApJS..134..181J} {134, 181}

\bibitem[\protect\citeauthoryear{{Jorstad}, {Marscher}, {Mattox}, {Aller},
  {Aller}, {Wehrle}  \& {Bloom}}{{Jorstad} et~al.}{2001b}]{Jorstad2001b}
{Jorstad} S.~G.,  {Marscher} A.~P.,  {Mattox} J.~R.,  {Aller} M.~F.,  {Aller}
  H.~D.,  {Wehrle} A.~E.,   {Bloom} S.~D.,  2001b, \mn@doi [\apj]
  {10.1086/321605}, \href
  {https://ui.adsabs.harvard.edu/abs/2001ApJ...556..738J} {556, 738}

\bibitem[\protect\citeauthoryear{{Jorstad} et~al.,}{{Jorstad}
  et~al.}{2005}]{Jorstad2005}
{Jorstad} S.~G.,  et~al., 2005, \mn@doi [\aj] {10.1086/444593}, \href
  {https://ui.adsabs.harvard.edu/abs/2005AJ....130.1418J} {130, 1418}

\bibitem[\protect\citeauthoryear{{Jorstad} et~al.,}{{Jorstad}
  et~al.}{2013}]{Jorstad2013}
{Jorstad} S.~G.,  et~al., 2013, \mn@doi [\apj] {10.1088/0004-637X/773/2/147},
  \href {https://ui.adsabs.harvard.edu/abs/2013ApJ...773..147J} {773, 147}

\bibitem[\protect\citeauthoryear{{Jorstad} et~al.,}{{Jorstad}
  et~al.}{2017}]{Jorstad2017}
{Jorstad} S.~G.,  et~al., 2017, \mn@doi [\apj] {10.3847/1538-4357/aa8407},
  \href {https://ui.adsabs.harvard.edu/abs/2017ApJ...846...98J} {846, 98}

\bibitem[\protect\citeauthoryear{{Kaspi}, {Smith}, {Netzer}, {Maoz}, {Jannuzi}
  \& {Giveon}}{{Kaspi} et~al.}{2000}]{Kaspi2000}
{Kaspi} S.,  {Smith} P.~S.,  {Netzer} H.,  {Maoz} D.,  {Jannuzi} B.~T.,
  {Giveon} U.,  2000, \mn@doi [\apj] {10.1086/308704}, \href
  {https://ui.adsabs.harvard.edu/abs/2000ApJ...533..631K} {533, 631}

\bibitem[\protect\citeauthoryear{{Kass} \& {Raftery}}{{Kass} \&
  {Raftery}}{1995}]{Kass1995}
{Kass} R E.,  {Raftery} A.~E.,  1995, Journal of the American Statistical
  Association, 90, 773

\bibitem[\protect\citeauthoryear{{Kellermann}}{{Kellermann}}{2003}]{Kellermann2003}
{Kellermann} K.~I.,  2003, in {Zensus} J.~A.,  {Cohen} M.~H.,   {Ros} E.,  eds,
   Astronomical Society of the Pacific Conference Series Vol. 300, Radio
  Astronomy at the Fringe. p.~185 (\mn@eprint {arXiv} {astro-ph/0304165})

\bibitem[\protect\citeauthoryear{{Kellermann} \& {Pauliny-Toth}}{{Kellermann}
  \& {Pauliny-Toth}}{1969}]{Kellermann1969}
{Kellermann} K.~I.,  {Pauliny-Toth} I.~I.~K.,  1969, \mn@doi [\apjl]
  {10.1086/180305}, \href
  {https://ui.adsabs.harvard.edu/abs/1969ApJ...155L..71K} {155, L71}

\bibitem[\protect\citeauthoryear{{Kellermann} et~al.,}{{Kellermann}
  et~al.}{2004}]{Kellermann2004}
{Kellermann} K.~I.,  et~al., 2004, \mn@doi [\apj] {10.1086/421289}, \href
  {https://ui.adsabs.harvard.edu/abs/2004ApJ...609..539K} {609, 539}

\bibitem[\protect\citeauthoryear{{Kellermann} et~al.,}{{Kellermann}
  et~al.}{2007}]{Kellermann2007}
{Kellermann} K.~I.,  et~al., 2007, \mn@doi [\apss] {10.1007/s10509-007-9622-5},
  \href {https://ui.adsabs.harvard.edu/abs/2007Ap&SS.311..231K} {311, 231}

\bibitem[\protect\citeauthoryear{{Komatsu} et~al.,}{{Komatsu}
  et~al.}{2011}]{Komatsu2011}
{Komatsu} E.,  et~al., 2011, \mn@doi [\apjs] {10.1088/0067-0049/192/2/18},
  \href {https://ui.adsabs.harvard.edu/abs/2011ApJS..192...18K} {192, 18}

\bibitem[\protect\citeauthoryear{{Liodakis}, {Hovatta}, {Huppenkothen},
  {Kiehlmann}, {Max-Moerbeck}  \& {Readhead}}{{Liodakis}
  et~al.}{2018}]{Liodakis2018}
{Liodakis} I.,  {Hovatta} T.,  {Huppenkothen} D.,  {Kiehlmann} S.,
  {Max-Moerbeck} W.,   {Readhead} A. C.~S.,  2018, \mn@doi [\apj]
  {10.3847/1538-4357/aae2b7}, \href
  {https://ui.adsabs.harvard.edu/abs/2018ApJ...866..137L} {866, 137}

\bibitem[\protect\citeauthoryear{{Lister} et~al.,}{{Lister}
  et~al.}{2009}]{Lister2009}
{Lister} M.~L.,  et~al., 2009, \mn@doi [\aj] {10.1088/0004-6256/138/6/1874},
  \href {https://ui.adsabs.harvard.edu/abs/2009AJ....138.1874L} {138, 1874}

\bibitem[\protect\citeauthoryear{{Lister} et~al.,}{{Lister}
  et~al.}{2013}]{Lister2013}
{Lister} M.~L.,  et~al., 2013, \mn@doi [\aj] {10.1088/0004-6256/146/5/120},
  \href {https://ui.adsabs.harvard.edu/abs/2013AJ....146..120L} {146, 120}

\bibitem[\protect\citeauthoryear{{Lister}, {Aller}, {Aller}, {Hovatta},
  {Max-Moerbeck}, {Readhead}, {Richards}  \& {Ros}}{{Lister}
  et~al.}{2015}]{Lister2015}
{Lister} M.~L.,  {Aller} M.~F.,  {Aller} H.~D.,  {Hovatta} T.,  {Max-Moerbeck}
  W.,  {Readhead} A.~C.~S.,  {Richards} J.~L.,   {Ros} E.,  2015, \mn@doi
  [\apjl] {10.1088/2041-8205/810/1/L9}, \href
  {https://ui.adsabs.harvard.edu/abs/2015ApJ...810L...9L} {810, L9}

\bibitem[\protect\citeauthoryear{{Lister}, {Aller}, {Aller}, {Hodge}, {Homan},
  {Kovalev}, {Pushkarev}  \& {Savolainen}}{{Lister} et~al.}{2018}]{Lister2018}
{Lister} M.~L.,  {Aller} M.~F.,  {Aller} H.~D.,  {Hodge} M.~A.,  {Homan} D.~C.,
   {Kovalev} Y.~Y.,  {Pushkarev} A.~B.,   {Savolainen} T.,  2018, \mn@doi
  [\apjs] {10.3847/1538-4365/aa9c44}, \href
  {https://ui.adsabs.harvard.edu/abs/2018ApJS..234...12L} {234, 12}

\bibitem[\protect\citeauthoryear{{Lister} et~al.,}{{Lister}
  et~al.}{2019}]{Lister2019}
{Lister} M.~L.,  et~al., 2019, \mn@doi [\apj] {10.3847/1538-4357/ab08ee}, \href
  {https://ui.adsabs.harvard.edu/abs/2019ApJ...874...43L} {874, 43}

\bibitem[\protect\citeauthoryear{{Lister}, {Homan}, {Kellermann}, {Kovalev},
  {Pushkarev}, {Ros}  \& {Savolainen}}{{Lister} et~al.}{2021}]{Lister2021}
{Lister} M.~L.,  {Homan} D.~C.,  {Kellermann} K.~I.,  {Kovalev} Y.~Y.,
  {Pushkarev} A.~B.,  {Ros} E.,   {Savolainen} T.,  2021, \mn@doi [\apj]
  {10.3847/1538-4357/ac230f}, \href
  {https://ui.adsabs.harvard.edu/abs/2021ApJ...923...30L} {923, 30}

\bibitem[\protect\citeauthoryear{{Marscher} et~al.,}{{Marscher}
  et~al.}{2010}]{Marscher2010}
{Marscher} A.~P.,  et~al., 2010, \mn@doi [\apjl]
  {10.1088/2041-8205/710/2/L126}, \href
  {https://ui.adsabs.harvard.edu/abs/2010ApJ...710L.126M} {710, L126}

\bibitem[\protect\citeauthoryear{{McLure} \& {Dunlop}}{{McLure} \&
  {Dunlop}}{2001}]{McLure2001}
{McLure} R.~J.,  {Dunlop} J.~S.,  2001, \mn@doi [\mnras]
  {10.1046/j.1365-8711.2001.04709.x}, \href
  {https://ui.adsabs.harvard.edu/abs/2001MNRAS.327..199M} {327, 199}

\bibitem[\protect\citeauthoryear{{McLure} \& {Dunlop}}{{McLure} \&
  {Dunlop}}{2004}]{McLure2004}
{McLure} R.~J.,  {Dunlop} J.~S.,  2004, \mn@doi [\mnras]
  {10.1111/j.1365-2966.2004.08034.x}, \href
  {https://ui.adsabs.harvard.edu/abs/2004MNRAS.352.1390M} {352, 1390}

\bibitem[\protect\citeauthoryear{{Meyer}, {Fossati}, {Georganopoulos}  \&
  {Lister}}{{Meyer} et~al.}{2011}]{Meyer2011}
{Meyer} E.~T.,  {Fossati} G.,  {Georganopoulos} M.,   {Lister} M.~L.,  2011,
  \mn@doi [\apj] {10.1088/0004-637X/740/2/98}, \href
  {https://ui.adsabs.harvard.edu/abs/2011ApJ...740...98M} {740, 98}

\bibitem[\protect\citeauthoryear{{Nolan} et~al.,}{{Nolan}
  et~al.}{2012}]{Nolan2012}
{Nolan} P.~L.,  et~al., 2012, \mn@doi [\apjs] {10.1088/0067-0049/199/2/31},
  \href {https://ui.adsabs.harvard.edu/abs/2012ApJS..199...31N} {199, 31}

\bibitem[\protect\citeauthoryear{{Paliya}, {Dom{\'\i}nguez}, {Ajello},
  {Olmo-Garc{\'\i}a}  \& {Hartmann}}{{Paliya} et~al.}{2021}]{Paliya2021}
{Paliya} V.~S.,  {Dom{\'\i}nguez} A.,  {Ajello} M.,  {Olmo-Garc{\'\i}a} A.,
  {Hartmann} D.,  2021, \mn@doi [\apjs] {10.3847/1538-4365/abe135}, \href
  {https://ui.adsabs.harvard.edu/abs/2021ApJS..253...46P} {253, 46}

\bibitem[\protect\citeauthoryear{{Pei}, {Fan}, {Liu}, {Yuan}, {Cai}, {Xiao},
  {Lin}  \& {Yang}}{{Pei} et~al.}{2016}]{Pei2016}
{Pei} Z.-Y.,  {Fan} J.-H.,  {Liu} Y.,  {Yuan} Y.-H.,  {Cai} W.,  {Xiao} H.-B.,
  {Lin} C.,   {Yang} J.-H.,  2016, \mn@doi [\apss] {10.1007/s10509-016-2822-0},
  \href {https://ui.adsabs.harvard.edu/abs/2016Ap&SS.361..237P} {361, 237}

\bibitem[\protect\citeauthoryear{{Pei}, {Fan}, {Bastieri}, {Sawangwit}  \&
  {Yang}}{{Pei} et~al.}{2019}]{Pei2019}
{Pei} Z.-Y.,  {Fan} J.-H.,  {Bastieri} D.,  {Sawangwit} U.,   {Yang} J.-H.,
  2019, \mn@doi [Research in Astronomy and Astrophysics]
  {10.1088/1674-4527/19/5/70}, \href
  {https://ui.adsabs.harvard.edu/abs/2019RAA....19...70P} {19, 070}

\bibitem[\protect\citeauthoryear{{Pei}, {Fan}, {Bastieri}, {Yang}  \&
  {Xiao}}{{Pei} et~al.}{2020}]{Pei2020}
{Pei} Z.,  {Fan} J.,  {Bastieri} D.,  {Yang} J.,   {Xiao} H.,  2020, \mn@doi
  [Science China Physics, Mechanics, and Astronomy]
  {10.1007/s11433-019-1454-6}, \href
  {https://ui.adsabs.harvard.edu/abs/2020SCPMA..6359511P} {63, 259511}

\bibitem[\protect\citeauthoryear{{Peterson} et~al.,}{{Peterson}
  et~al.}{1999}]{Peterson1999}
{Peterson} B.~M.,  et~al., 1999, \mn@doi [\apj] {10.1086/306604}, \href
  {https://ui.adsabs.harvard.edu/abs/1999ApJ...510..659P} {510, 659}

\bibitem[\protect\citeauthoryear{{Peterson} et~al.,}{{Peterson}
  et~al.}{2000}]{Peterson2000}
{Peterson} B.~M.,  et~al., 2000, \mn@doi [\apj] {10.1086/309518}, \href
  {https://ui.adsabs.harvard.edu/abs/2000ApJ...542..161P} {542, 161}

\bibitem[\protect\citeauthoryear{{Piner} \& {Edwards}}{{Piner} \&
  {Edwards}}{2014}]{Piner2014}
{Piner} B.~G.,  {Edwards} P.~G.,  2014, \mn@doi [\apj]
  {10.1088/0004-637X/797/1/25}, \href
  {https://ui.adsabs.harvard.edu/abs/2014ApJ...797...25P} {797, 25}

\bibitem[\protect\citeauthoryear{{Piner} \& {Edwards}}{{Piner} \&
  {Edwards}}{2018}]{Piner2018}
{Piner} B.~G.,  {Edwards} P.~G.,  2018, \mn@doi [\apj]
  {10.3847/1538-4357/aaa425}, \href
  {https://ui.adsabs.harvard.edu/abs/2018ApJ...853...68P} {853, 68}

\bibitem[\protect\citeauthoryear{{Piner}, {Mahmud}, {Fey}  \&
  {Gospodinova}}{{Piner} et~al.}{2007}]{Piner2007}
{Piner} B.~G.,  {Mahmud} M.,  {Fey} A.~L.,   {Gospodinova} K.,  2007, \mn@doi
  [\aj] {10.1086/514812}, \href
  {https://ui.adsabs.harvard.edu/abs/2007AJ....133.2357P} {133, 2357}

\bibitem[\protect\citeauthoryear{{Piner} et~al.,}{{Piner}
  et~al.}{2012}]{Piner2012}
{Piner} B.~G.,  et~al., 2012, \mn@doi [\apj] {10.1088/0004-637X/758/2/84},
  \href {https://ui.adsabs.harvard.edu/abs/2012ApJ...758...84P} {758, 84}

\bibitem[\protect\citeauthoryear{{Rawlings} \& {Saunders}}{{Rawlings} \&
  {Saunders}}{1991}]{Rawlings1991}
{Rawlings} S.,  {Saunders} R.,  1991, \mn@doi [\nat] {10.1038/349138a0}, \href
  {https://ui.adsabs.harvard.edu/abs/1991Natur.349..138R} {349, 138}

\bibitem[\protect\citeauthoryear{{Readhead}}{{Readhead}}{1994}]{Readhead1994}
{Readhead} A. C.~S.,  1994, \mn@doi [\apj] {10.1086/174038}, \href
  {https://ui.adsabs.harvard.edu/abs/1994ApJ...426...51R} {426, 51}

\bibitem[\protect\citeauthoryear{{Rees}}{{Rees}}{1966}]{Rees1966}
{Rees} M.~J.,  1966, \mn@doi [\nat] {10.1038/211468a0}, \href
  {https://ui.adsabs.harvard.edu/abs/1966Natur.211..468R} {211, 468}

\bibitem[\protect\citeauthoryear{{Scarpa} \& {Falomo}}{{Scarpa} \&
  {Falomo}}{1997}]{Scarpa1997}
{Scarpa} R.,  {Falomo} R.,  1997, \aap, \href
  {https://ui.adsabs.harvard.edu/abs/1997A&A...325..109S} {325, 109}

\bibitem[\protect\citeauthoryear{{Shaw} et~al.,}{{Shaw}
  et~al.}{2012}]{Shaw2012}
{Shaw} M.~S.,  et~al., 2012, \mn@doi [\apj] {10.1088/0004-637X/748/1/49}, \href
  {https://ui.adsabs.harvard.edu/abs/2012ApJ...748...49S} {748, 49}

\bibitem[\protect\citeauthoryear{{Shen} et~al.,}{{Shen}
  et~al.}{2011}]{Shen2011}
{Shen} Y.,  et~al., 2011, \mn@doi [\apjs] {10.1088/0067-0049/194/2/45}, \href
  {https://ui.adsabs.harvard.edu/abs/2011ApJS..194...45S} {194, 45}

\bibitem[\protect\citeauthoryear{{Sikora}, {Begelman}  \& {Rees}}{{Sikora}
  et~al.}{1994}]{Sikora1994}
{Sikora} M.,  {Begelman} M.~C.,   {Rees} M.~J.,  1994, \mn@doi [\apj]
  {10.1086/173633}, \href
  {https://ui.adsabs.harvard.edu/abs/1994ApJ...421..153S} {421, 153}

\bibitem[\protect\citeauthoryear{{Sokolov} \& {Marscher}}{{Sokolov} \&
  {Marscher}}{2005}]{Sokolov2005}
{Sokolov} A.,  {Marscher} A.~P.,  2005, \mn@doi [\apj] {10.1086/431321}, \href
  {https://ui.adsabs.harvard.edu/abs/2005ApJ...629...52S} {629, 52}

\bibitem[\protect\citeauthoryear{{Tan}, {Xue}, {Du}, {Xi}, {Wang}  \&
  {Xie}}{{Tan} et~al.}{2020}]{Tan2020}
{Tan} C.,  {Xue} R.,  {Du} L.-M.,  {Xi} S.-Q.,  {Wang} Z.-R.,   {Xie} Z.-H.,
  2020, \mn@doi [\apjs] {10.3847/1538-4365/ab8cc6}, \href
  {https://ui.adsabs.harvard.edu/abs/2020ApJS..248...27T} {248, 27}

\bibitem[\protect\citeauthoryear{{Tanaka} et~al.,}{{Tanaka}
  et~al.}{2015}]{Tanaka2015}
{Tanaka} Y.~T.,  et~al., 2015, \mn@doi [\apjl] {10.1088/2041-8205/799/2/L18},
  \href {https://ui.adsabs.harvard.edu/abs/2015ApJ...799L..18T} {799, L18}

\bibitem[\protect\citeauthoryear{{Taylor}, {Vermeulen}, {Readhead}, {Pearson},
  {Henstock}  \& {Wilkinson}}{{Taylor} et~al.}{1996}]{Taylor1996}
{Taylor} G.~B.,  {Vermeulen} R.~C.,  {Readhead} A.~C.~S.,  {Pearson} T.~J.,
  {Henstock} D.~R.,   {Wilkinson} P.~N.,  1996, \mn@doi [\apjs]
  {10.1086/192354}, \href
  {https://ui.adsabs.harvard.edu/abs/1996ApJS..107...37T} {107, 37}

\bibitem[\protect\citeauthoryear{{Urry} \& {Padovani}}{{Urry} \&
  {Padovani}}{1995}]{Urry1995}
{Urry} C.~M.,  {Padovani} P.,  1995, \mn@doi [\pasp] {10.1086/133630}, \href
  {https://ui.adsabs.harvard.edu/abs/1995PASP..107..803U} {107, 803}

\bibitem[\protect\citeauthoryear{{Vermeulen}, {Readhead}  \&
  {Backer}}{{Vermeulen} et~al.}{1994}]{Vermeulen1994}
{Vermeulen} R.~C.,  {Readhead} A.~C.~S.,   {Backer} D.~C.,  1994, \mn@doi
  [\apjl] {10.1086/187433}, \href
  {https://ui.adsabs.harvard.edu/abs/1994ApJ...430L..41V} {430, L41}

\bibitem[\protect\citeauthoryear{{Vestergaard}}{{Vestergaard}}{2002}]{Vestergaard2002}
{Vestergaard} M.,  2002, \mn@doi [\apj] {10.1086/340045}, \href
  {https://ui.adsabs.harvard.edu/abs/2002ApJ...571..733V} {571, 733}

\bibitem[\protect\citeauthoryear{{Vestergaard} \& {Peterson}}{{Vestergaard} \&
  {Peterson}}{2006}]{Vestergaard2006}
{Vestergaard} M.,  {Peterson} B.~M.,  2006, \mn@doi [\apj] {10.1086/500572},
  \href {https://ui.adsabs.harvard.edu/abs/2006ApJ...641..689V} {641, 689}

\bibitem[\protect\citeauthoryear{{Villata} et~al.,}{{Villata}
  et~al.}{2006}]{Villata2006}
{Villata} M.,  et~al., 2006, \mn@doi [\aap] {10.1051/0004-6361:20064817}, \href
  {https://ui.adsabs.harvard.edu/abs/2006A&A...453..817V} {453, 817}

\bibitem[\protect\citeauthoryear{{Weaver} et~al.,}{{Weaver}
  et~al.}{2022}]{Weaver2022}
{Weaver} Z.~R.,  et~al., 2022, \mn@doi [\apjs] {10.3847/1538-4365/ac589c},
  \href {https://ui.adsabs.harvard.edu/abs/2022ApJS..260...12W} {260, 12}

\bibitem[\protect\citeauthoryear{{Willott}, {Rawlings}, {Blundell}  \&
  {Lacy}}{{Willott} et~al.}{1999}]{Willott1999}
{Willott} C.~J.,  {Rawlings} S.,  {Blundell} K.~M.,   {Lacy} M.,  1999, \mn@doi
  [\mnras] {10.1046/j.1365-8711.1999.02907.x}, \href
  {https://ui.adsabs.harvard.edu/abs/1999MNRAS.309.1017W} {309, 1017}

\bibitem[\protect\citeauthoryear{{Wills}, {Wills}, {Breger}, {Antonucci}  \&
  {Barvainis}}{{Wills} et~al.}{1992}]{Wills1992}
{Wills} B.~J.,  {Wills} D.,  {Breger} M.,  {Antonucci} R.~R.~J.,   {Barvainis}
  R.,  1992, \mn@doi [\apj] {10.1086/171869}, \href
  {https://ui.adsabs.harvard.edu/abs/1992ApJ...398..454W} {398, 454}

\bibitem[\protect\citeauthoryear{{Xiao}, {Pei}, {Xie}, {Hao}, {Yang}, {Yuan},
  {Liu}  \& {Fan}}{{Xiao} et~al.}{2015}]{Xiao2015}
{Xiao} H.~B.,  {Pei} Z.~Y.,  {Xie} H.~J.,  {Hao} J.~M.,  {Yang} J.~H.,  {Yuan}
  Y.~H.,  {Liu} Y.,   {Fan} J.~H.,  2015, \mn@doi [\apss]
  {10.1007/s10509-015-2433-1}, \href
  {https://ui.adsabs.harvard.edu/abs/2015Ap&SS.359...39X} {359, 39}

\bibitem[\protect\citeauthoryear{{Xiao} et~al.,}{{Xiao}
  et~al.}{2019}]{Xiao2019}
{Xiao} H.,  et~al., 2019, \mn@doi [Science China Physics, Mechanics, and
  Astronomy] {10.1007/s11433-018-9371-x}, \href
  {https://ui.adsabs.harvard.edu/abs/2019SCPMA..62l9811X} {62, 129811}

\bibitem[\protect\citeauthoryear{{Xiao}, {Fan}, {Rando}, {Zhu}  \& {Hu}}{{Xiao}
  et~al.}{2020}]{Xiao2020}
{Xiao} H.,  {Fan} J.,  {Rando} R.,  {Zhu} J.,   {Hu} L.,  2020, \mn@doi
  [Astronomische Nachrichten] {10.1002/asna.202013733}, \href
  {https://ui.adsabs.harvard.edu/abs/2020AN....341..462X} {341, 462}

\bibitem[\protect\citeauthoryear{{Xiao}, {Ouyang}, {Zhang}, {Fu}, {Zhang},
  {Zeng}  \& {Fan}}{{Xiao} et~al.}{2022}]{Xiao2022a}
{Xiao} H.,  {Ouyang} Z.,  {Zhang} L.,  {Fu} L.,  {Zhang} S.,  {Zeng} X.,
  {Fan} J.,  2022, \mn@doi [\apj] {10.3847/1538-4357/ac36da}, \href
  {https://ui.adsabs.harvard.edu/abs/2022ApJ...925...40X} {925, 40}

\bibitem[\protect\citeauthoryear{{Xie}, {Liu}, {Liu}, {Lu}, {Li}  \&
  {Zhu}}{{Xie} et~al.}{1991}]{Xie1991}
{Xie} G.~Z.,  {Liu} F.~K.,  {Liu} B.~F.,  {Lu} R.~W.,  {Li} K.~H.,   {Zhu}
  Y.~Y.,  1991, \aap, \href
  {https://ui.adsabs.harvard.edu/abs/1991A&A...249...65X} {249, 65}

\bibitem[\protect\citeauthoryear{{Xiong} \& {Zhang}}{{Xiong} \&
  {Zhang}}{2014}]{Xiong2014}
{Xiong} D.~R.,  {Zhang} X.,  2014, \mn@doi [\mnras] {10.1093/mnras/stu755},
  \href {https://ui.adsabs.harvard.edu/abs/2014MNRAS.441.3375X} {441, 3375}

\bibitem[\protect\citeauthoryear{{Yang}, {Xiao}, {Wang}, {Yang}, {Pei}, {Wu},
  Yu-Hai  \& {Fan}}{{Yang} et~al.}{2022}]{Yang2022}
{Yang} W.-X.,  {Xiao} H.-B.,  {Wang} H.-G.,  {Yang} J.-H.,  {Pei} Z.-Y.,  {Wu}
  D.-X.,  Yu-Hai Y.,   {Fan} J.,  2022, Research in Astronomy and Astrophysics

\bibitem[\protect\citeauthoryear{{Zhang} \& {Fan}}{{Zhang} \&
  {Fan}}{2008}]{Zhang2008}
{Zhang} Y.-W.,  {Fan} J.-H.,  2008, \mn@doi [\cjaa] {10.1088/1009-9271/8/4/02},
  \href {https://ui.adsabs.harvard.edu/abs/2008ChJAA...8..385Z} {8, 385}

\bibitem[\protect\citeauthoryear{{Zhang}, {Liang}, {Zhang}  \& {Bai}}{{Zhang}
  et~al.}{2012}]{Zhang2012}
{Zhang} J.,  {Liang} E.-W.,  {Zhang} S.-N.,   {Bai} J.~M.,  2012, \mn@doi
  [\apj] {10.1088/0004-637X/752/2/157}, \href
  {https://ui.adsabs.harvard.edu/abs/2012ApJ...752..157Z} {752, 157}

\bibitem[\protect\citeauthoryear{{Zhang}, {Chen}, {Xiao}, {Cai}  \&
  {Fan}}{{Zhang} et~al.}{2020}]{Zhang2020}
{Zhang} L.,  {Chen} S.,  {Xiao} H.,  {Cai} J.,   {Fan} J.,  2020, \mn@doi
  [\apj] {10.3847/1538-4357/ab9180}, \href
  {https://ui.adsabs.harvard.edu/abs/2020ApJ...897...10Z} {897, 10}

\makeatother
\end{thebibliography}




\bsp	
\label{lastpage}
\end{document}